\documentclass[useAMS,usenatbib]{mn2e}
\usepackage{epsfig}
\usepackage{color} 
\usepackage{natbib}
\usepackage{deluxetable}
\usepackage{amssymb}
\usepackage{rotating}
\usepackage{graphicx}
\usepackage{caption}
\usepackage{longtable}
\usepackage{multicol}
\usepackage{booktabs}
\usepackage[colorlinks=true,citecolor=blue]{hyperref}
\usepackage{color}
\usepackage{graphicx}
\usepackage{tikz}
\usetikzlibrary{tikzmark}
\usepackage{amsmath, amsfonts, epsfig, xspace}
\def\dndz{$dN/dz$~}
\def\ew{W$_{r}$}

\def\mgi{Mg~{\sc i}~} 
\def\mgii{Mg~{\sc ii}~}

\def\feii{Fe~{\sc ii}~}

\def\znii{Zn~{\sc ii}~}
\def\mnii{Mn~{\sc ii}~}
\def\aliii{Al~{\sc iii}~} 

\def\civ{C~{\sc iv}~}

\def\siiv{Si~{\sc ii}~}
 
\title[Mg~{\sc ii} number density towards blazar]{On the incidence of Mg~{\sc ii} absorbers along the blazar sightlines}
\author[$Mishra$ et al.]{{\Large S. Mishra$^{1}$\thanks{E-mail: sapna@aries.res.in(SM)},
H. Chand$^{1}$, Gopal-Krishna$^{2}$\thanks{Platinum Jubilee Senior Scientist, The National Academy of Sciences, India.}, R. Joshi$^{3,4}$, Y. A. Shchekinov$^{5,6}$, T. A. Fatkhullin$^{7}$}\\\\
$^{1}$Aryabhatta Research Institute of Observational Sciences (ARIES), Manora Peak, Nainital $-$ 263002, India\\
  $^{2}$UM-DAE Centre for Excellence in Basic Sciences, University of Mumbai, Mumbai 400098, India\\
  $^{3}$Inter-University Centre for Astronomy and Astrophysics, Post Bag 4, Ganeshkhind, Pune 411007, India\\
$^{4}$Kavli Institute for Astronomy and Astrophysics, Peking University, Beijing 100871, China\\
  $^{5}$Lebedev Physical Institute, Russian Academy of Sciences, 53 Leninsky Ave., Moscow 119991, Russia\\
  $^{6}$Raman Research Institute, Sadashiva Nagar, Bangalore 560080, India\\
$^{7}$Special Astrophysical Observatory, Russian Academy of Science, Karachai-Cherkessia, 369167, Russia\\
}

\begin{document}
\date{Accepted ---. Received ---; in original form ---}

\pagerange{\pageref{firstpage}--\pageref{lastpage}} \pubyear{2017}

\maketitle

\label{firstpage}

\begin{abstract}
 It is widely believed that the cool gas clouds traced by Mg~{\sc ii}
 absorption, within a velocity offset of 5000 kms$^{-1}$ relative to
 the background quasar are mostly associated with the quasar itself,
 whereas the absorbers seen at larger velocity offsets towards us are
 intervening absorber systems and hence their existence is completely
 independent of the background quasar. Recent evidence by
 \citet[][hereinafter BBM]{Bergeron2011A&A...525A..51B} has seriously
 questioned this paradigm, by showing that the number density of
 intervening Mg~{\sc ii} absorbers towards the 45 blazars in their
 sample is nearly 2 times the expectation based on the Mg~{\sc ii}
 absorption systems seen towards normal QSOs. Given its serious
 implications, it becomes important to revisit this finding, by
 enlarging the blazar sample and subjecting it to an independent
 analysis. Here, we first report the outcome of our re-analysis of the
 available spectroscopic data for the BBM sample itself. Our analysis
 of the BBM sample reproduces their claimed factor of 2 excess of
 $dN/dz$ along blazar sightlines, vis-a-vis normal QSOs. We have also
 assembled a $\sim$3 times larger sample of blazars, albeit with
 moderately sensitive optical spectra. Using this sample
 together with the BBM sample, our analysis shows that the $dN/dz$ of
 the Mg~{\sc ii} absorbers statistically matches that known for normal
 QSO sightlines.  Further, the analysis indicates that associated
 absorbers might be contributing significantly to the estimated
 $dN/dz$ upto offset speeds $\Delta v \sim 0.2c$ relative to the blazar.
\end{abstract}


\section{Introduction}
\label{sec:intro_mgiidndz}
Analysis of the narrow absorption-line systems (in the spectra of
quasars) has emerged as a powerful probe of the physical conditions of
the gaseous medium of intervening galaxies, particularly when they lie
at extremely large distances and hence too faint for direct
imaging/spectroscopy even with the largest telescopes
\citep{Bahcall1966ApJ...144..847B,Wolfe2005ARA&A..43..861W,Kulkarni2012ApJ...749..176K}. It
is widely held that the cool gas clouds (e.g., Mg~{\sc ii} absorption
systems) with velocities offsets $\beta$c up to $\sim$ 5000 km
s$^{-1}$ relative to the background quasar are gravitationally bound
to the quasar itself \citep[i.e., `associated systems', see][and
  references
  therein]{Anderson1987AJ.....94..278A,Pushpa1989ApJ...347..627K,1994A&A...287..719M},
whereas absorbers showing larger velocity offsets directed towards us are
intervening systems probably associated with foreground galaxies and,
consequently, their existence should be totally independent of the
background quasar.  A few recent studies, however, seem to question
this canonical view, based on the differing estimates for the
incidence rates of intervening systems (like Mg~{\sc ii} absorption
having $\beta c \ge 5000$ km s$^{-1}$) detected towards different
types of background sources, such as normal QSOs, gamma-ray bursters
(GRBs) and blazars \citep[][hereinafter
  BBM]{Stoke1997ApJ...489L..17S,Prochter2006ApJ...648L..93P,Sudilovsky2007ApJ...669..741S,
  Vergani2009A&A...503..771V,Tejos2009ApJ...706.1309T,Bergeron2011A&A...525A..51B}.
It has been also claimed by BBM
and \citet{2009ApJ...697..345C} that associated systems having a
significantly relativistic speed relative to the quasar may also be
present (BBM), e.g., when the quasar is undergoing powerful jet
activity and/or ejecting high speed accretion-disk outflows.  A
possible way to differentiate between these possibilities would be to
check if the incidence rates, $dN/dz$, of intervening absorbers differ
depending on whether the background sources are non-blazars, or
blazars whose powerful relativistic jets are therefore expected to be
pointed close to our direction and, consequently the jet-accelerated
potential absorbers would lie along the line-of-sight. Indeed, this
expectation is echoed in the unexpected finding of BBM that the
$dN/dz$ of Mg~{\sc ii} absorption systems (for strong absorbers having
a rest-frame equivalent width W$_{r} ~ \geq$ 1~\AA ) towards
blazars is $\sim$ 2 times larger (at 3$\sigma$ confidence) than the
value established for the sightlines to normal quasars (QSOs). An
even greater excess had earlier been reported by Stocke \& Rector
(1997), albeit using a much smaller sample of blazars. On the other
hand, a recent analysis by \citet{Chand2012ApJ...754...38C} of the
existing high-resolution spectra of a sample of about 115
flat-spectrum radio-loud quasars (FSRQs, of non-blazar type) did not
show any excess in the incidence of Mg~{\sc ii} absorption systems, as
compared to QSOs. They reconciled the two seemingly discrepant results
by appealing to the jet orientation scenario which lies at the heart
of the Unified Scheme for powerful extragalactic radio sources
\citep[e.g. see,][]{1993ARA&A..31..473A}. In their explanation, since
the jets in FSRQs are thought to be less closely aligned to the
line-of-sight, any gas clouds accelerated outward by the powerful jets
are unlikely to appear in the foreground of the quasar's nucleus and
hence escape being detected in absorption against its
bright optical emission.  Later, \citet{Joshi2013MNRAS.435..346J}
extended this probe by analyzing a large set of redshift-matched
sightlines to 3975 radio core-dominated (CDQs, i.e., FSRQs) and 1583
radio lobe-dominated (LDQs) quasars. While, overall, only a marginal (9\% at
1.5$\sigma$ significance) excess of $dN/dz$ was found towards the
FSRQ sightlines, as compared to the sightlines to normal QSOs, they
showed that the excess becomes quite significant (3.75$\sigma$) when
the comparison is restricted to the absorbers having offset speeds
i.e., $\beta  < 0.1c$ relative to the background quasar.
 Similarly, \citet{2011ApJ...742...44T} have used observations of Fe XXV/XXVI
K-shell resonance lines in the X-ray band and found their outflow velocity
distribution spans from $\sim$ 10,000 kms$^{-1}$ up to $\sim$ 100,000
kms$^{-1}$ ($\sim$ 0.3c), with a peak and mean value of $\sim$ 42,000
kms$^{-1}$ ($\sim$ 0.14c), for highly ionised gas clouds with column
densities of N$_{H} \approx$ 10$^{23}$ cm$^{-2}$ located within the
central parsec of the AGN.\par
In this context, it is important to emphasize that even though BBM's analysis
has employed very high-sensitivity spectral data, their result rests
on just 45 blazars, due to which small number statistics might be at
work.  It is worthwhile recalling that the 4-fold excess of $dN/dz$
along the GRB sightlines, inferred by
\citet{Prochter2006ApJ...648L..93P} using just 14 GRBs, has
subsequently been pronounced as a possible statistical fluke, on the
basis of a 3 times larger set of sightlines
\citep{Cucchiara2013ApJ...773...82C}.  Given its potentially deep
ramifications, it is therefore desirable to revisit the BBM claim of
excess $dN/dz$ towards blazars, by enlarging the blazar sample and
carrying out an independent analysis. The present study is motivated
by this objective.  \\ The paper is organized as follows, in Section 2
we describe our sample, while in Section 3 our data analysis procedure
is outlined. The results are given in Section 4, followed by a brief
discussion and conclusions in Section 5.
\section{The Sample}
\label{sample}
\begin{table}
 \centering
 \begin{minipage}[10]{140mm}
\caption{The spectral data sourcing for our enlarged sample of blazars.}
\label{table:sample}
{\scriptsize
\begin{tabular}{@{}lll lll l @{}} 
\hline \hline
Archive    &   Instrument	&  Content                            & Resolution \\
\hline
ESO       &   FORS-1/2         &  48 blazars found (10 taken\footnote{Ten blazars do not have useful redshift path (i.e., not satisfying our\\ selection criteria-ii), another 28 lack emission redshift ($48-10-28=10$).})
                                                                       &  900 \\
ESO       &   X-SHOOTER         &  17 found(8 taken\footnote{Two blazars do not have useful redshift path (i.e., not satisfying the \\selection criteria-ii), another 7 lack emission redshift ($17-2-7=8$).})
                                                                       &   3600 \\
 ESO       &   UVES             &  1 found (taken)                     &  40000 \\
 SAO       &   SCORPIO       &  3 new observations (3 taken)            &  818   \\
 KECK      &   LRIS             &  2 (both taken)                      &  9800  \\
 SDSS      &   BOSS             &  622 found (196 taken\footnote{Excluded 69 sources with SNR $<$ 5 (i.e. not satisfying the selection \\criteria-i,  207 lack emission redshift, and 138 sources were excluded for \\not meeting the selection criteria-ii), 12 sources were excluded since their\\ spectra had been already taken from the other archives listed in this\\ table ($622-69-207-138-12=196$).})
                                                                       & 2500 \\
 BBM       &   FORS-1           &  42 (all 42 taken)                   & 900  \\
 \hline
\end{tabular}
}                                                            
\end{minipage}                                                           
\end{table}
The blazar sample employed in our analysis is an amalgamation of 3
sets of blazars extracted from the catalogues published by
\citet[][hereafter ROMA-BZCAT]{Massaro2009yCat..34950691M},
\citet[][hereafter VV]{Veron2010A&A...518A..10V}, and
\citet[][hereafter, Padovani-Catalogue]{Padovani1995MNRAS.277.1477P}.
From the ROMA-BZCAT we selected sources classified as BZB (implying
confirmed BL Lac), resulting in a set of 1059 blazars. From the VV
catalogue we selected the sources classified either as `BL' (i.e.,
confirmed BL Lac), or `HP' (a confirmed highly polarized quasar). This
resulted in a set of 729 confirmed blazars from this catalogue.
Accounting for the 480 sources that are common to these two sets, led
to a list of 1308 confirmed blazars.  The Padovani-catalogue also
classifies BL Lacs objects using homogeneous criteria. It contains a
total of 233 blazars, of which 189 were already in the above two sets
(among them 169 blazars of the Padovani-catalogue are in BZ-ROMA,
while 20 in the VV catalogue). Their exclusion left us with 44 blazars
solely contributed by the Padovani-catalogue.  Merging these 3 sets
resulted in our final `parent sample' of 1352 confirmed blazars.\par
We then performed an extensive search for optical spectra of our
parent sample of blazars, in the archives of the Sloan Digital Sky
Survey\footnote{https://dr12.sdss.org/bulkSpectra} (SDSS), the
European Southern
Observatory\footnote{http://archive.eso.org/eso/eso$\_$archive$\_$main.html}
(ESO) and the
KECK\footnote{https://koa.ipac.caltech.edu/cgi-bin/KOA/nph-KOAlogin}
Observatory. We applied two main selection filters: (i) median SNR of
the entire spectrum should be more than 5, so that false detections of
Mg~{\sc ii} line are minimized, and (ii) the blazar's redshift should
allow at least $10\times(1+z_{em}$)\AA~ wide coverage in the available
spectrum, of the region between the Ly$\alpha$ and Mg~{\sc ii}
emission lines; this would ensure that the observed spectrum can be
used to search for the Mg~{\sc ii} doublet due to at least one
absorber (given that the two components of the doublet Mg~{\sc ii}
$\lambda$ 2796, 2803, are separated by 8~\AA~ in the rest frame).\par
In the ESO archive, after excluding the spectra of the 42 BBM blazars,
which had been taken using the FOcal Reducer and the low-dispersion
Spectrograph (FORS1) at the ESO observatory, we found that for 66 of our
blazars (within 1 arcmin search radius) a spectrum with SNR $> 5$
was available either in the reduced form, from the ESO-advanced data
product \footnote{http://archive.eso.org/wdb/wdb/adp/phase3\_main/form}(17
observed using X-shooter spectrograph, 1 observed using the Ultraviolet
and the Visual Echelle Spectrograph (UVES) ), or we were at least able to
extract the spectra based on their raw images using associated
calibration files (48 observed using FORS-1,2).  Among these 66
blazars, emission redshift was available for only 31, out of which
just 16 were found useful for Mg~{\sc ii}~ absorber search after
applying our emission redshift constraint mentioned above. For the
 remaining 35 blazars with unknown emission redshifts we could
establish a lower redshift limit for 4 blazars using the redshift of
the observed most redshifted Mg~{\sc ii}~ absorption doublet. One of
 these 4 had to be excluded as the Mg~{\sc ii}~ derived
redshift was not yielding adequate redshift path (i.e., not satisfying
the selection criteria-ii). Here, it is also important to clarify that the spectral
region containing this most redshifted absorption doublet was excluded
for the purpose of computing $dN/dz$, in order to keep the estimate
free of bias resulting from exclusion of those blazars with unknown redshift,
for which even a lower limit to redshift could not be established (using the \mgii absorption doublet).
\par
For another three blazars from our `parent-sample' (see above), viz,
J145127$+$635426, J165248$+$363212, J182406$+$565100, we have newly
obtained spectra using the SCORPIO spectrograph (using VPHG1200 grism)
mounted on the 6-m telescope at the Special Astrophysical Observatory
(SAO). Inclusion of another 2 blazars viz, J001937$+$202146,
J043337$+$290555, became possible due to the availability of their
spectra in the KECK archive, taken with the Low Resolution Imaging
Spectrometer (LRIS).\par For 622 blazars in our `parent sample' we
could find spectra in the SDSS archives (within a tolerance of 2
arcsec) in reduced form, covering a wavelength range
3800-10000\AA. Excluding the 69 spectra with SNR $< 5$, left us with
good quality spectra for 553 blazars. Among these, emission redshifts
were available for 277 blazars, out of which 150 sources were found
useful for Mg~{\sc ii}~ absorber search, after meeting our
aforementioned emission redshift criterion (selection
criterion-ii). In addition, from among the 276 blazars with unknown
emission redshift, a lower redshift limit could be set for 59 sources,
using the detected Mg~{\sc ii}~ absorption feature. However, one of
these 59 sources (viz, J130008.5$+$175538) had to be excluded as it
did not meet our criterion of useful minimum redshift path (i.e., the
selection criterion-ii). In addition, after excluding another 12
blazars as they are already included in our above sample from other
resources (ESO archive and SAO observations) we are left with 196
blazars solely contributed by the SDSS, which are found satisfactory
for the purpose of our Mg~{\sc ii}~ absorption-line search.\par To
recapitulate, we have assembled a sample of 220 blazars (SDSS: 196,
ESO: 19, SAO: 3, KECK: 2) as summarized in Table~\ref{table:sample},
to make a search for intervening Mg~{\sc ii}~ absorbers. Out of these,
only a lower limit to $z_{em}$ is available for 58 blazars (i.e. 54
SDSS, 3 ESO and 1 KECK). We have dicussed three out of them in
Appendix~\ref{ap:appendix_A} where we have also shown their
represtative spectra as well. \par
Further, as
mentioned above, we have also made use of the BBM blazar sample to
revisit their conclusion (section 1), by subjecting it to an
independent data reduction and analysis procedure, as followed in the
present work. The sample employed in their analysis consists of 45
blazars. For 42 of them, we could obtain the raw spectral data from
the ESO
archive\footnote{http://archive.eso.org/eso/eso\_archive\_main.html}
based on their program ID 080:A-0276, 081:A-0193. The raw data used in
the BBM analysis for the remaining 3 (northern sky) blazars were not
accessible and hence they could not be included in our analysis.\par
\begin{figure*}
  \hspace{-0.6in}
  \begin{minipage}[]{0.4\textwidth}
  \includegraphics[width=0.8\textwidth,height=0.4\textheight,angle=90]{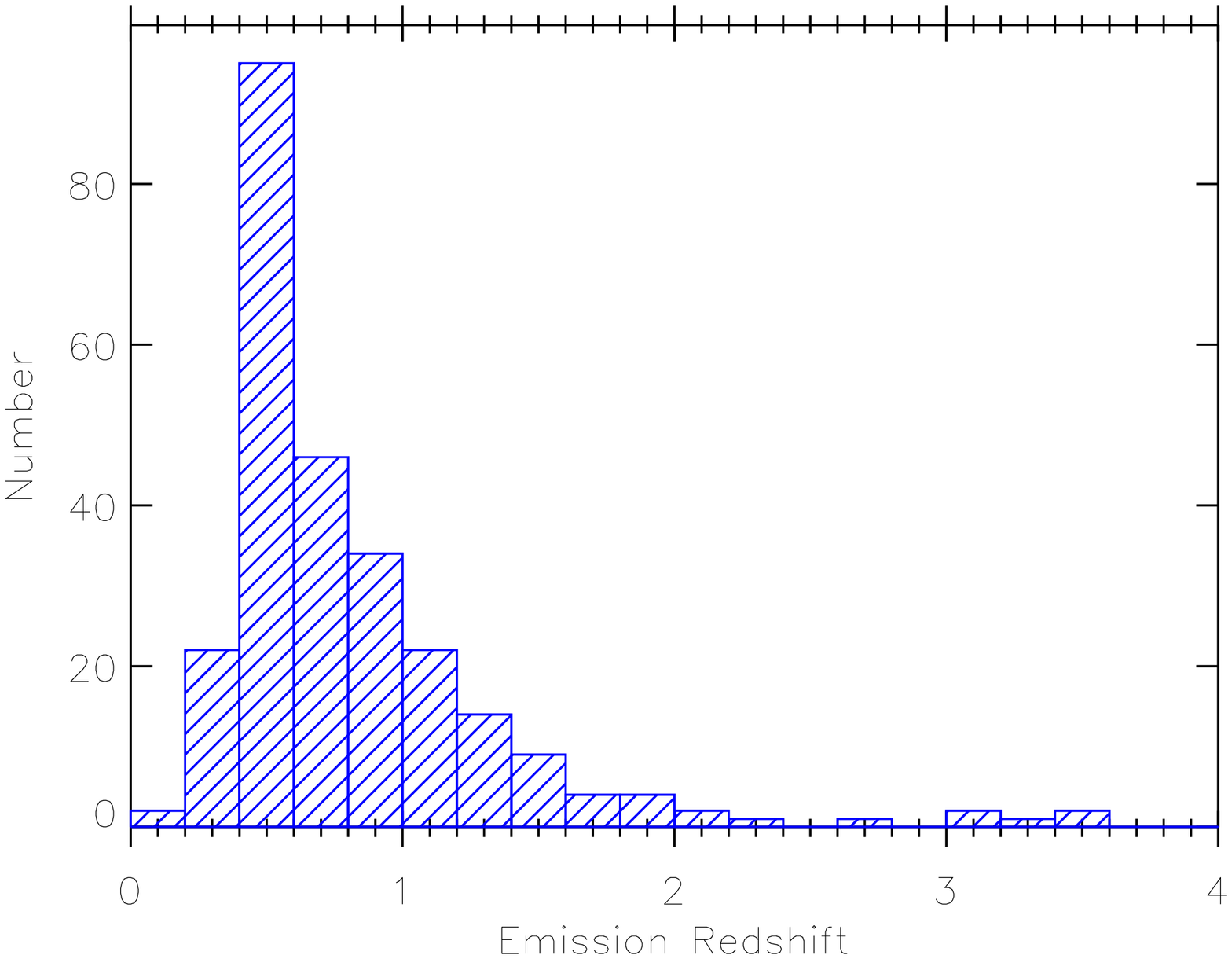}    
  \end{minipage}
  \hspace{0.6in}
  \begin{minipage}[]{0.4\textwidth}
  \includegraphics[width=0.8\textwidth,height=0.4\textheight,angle=90]{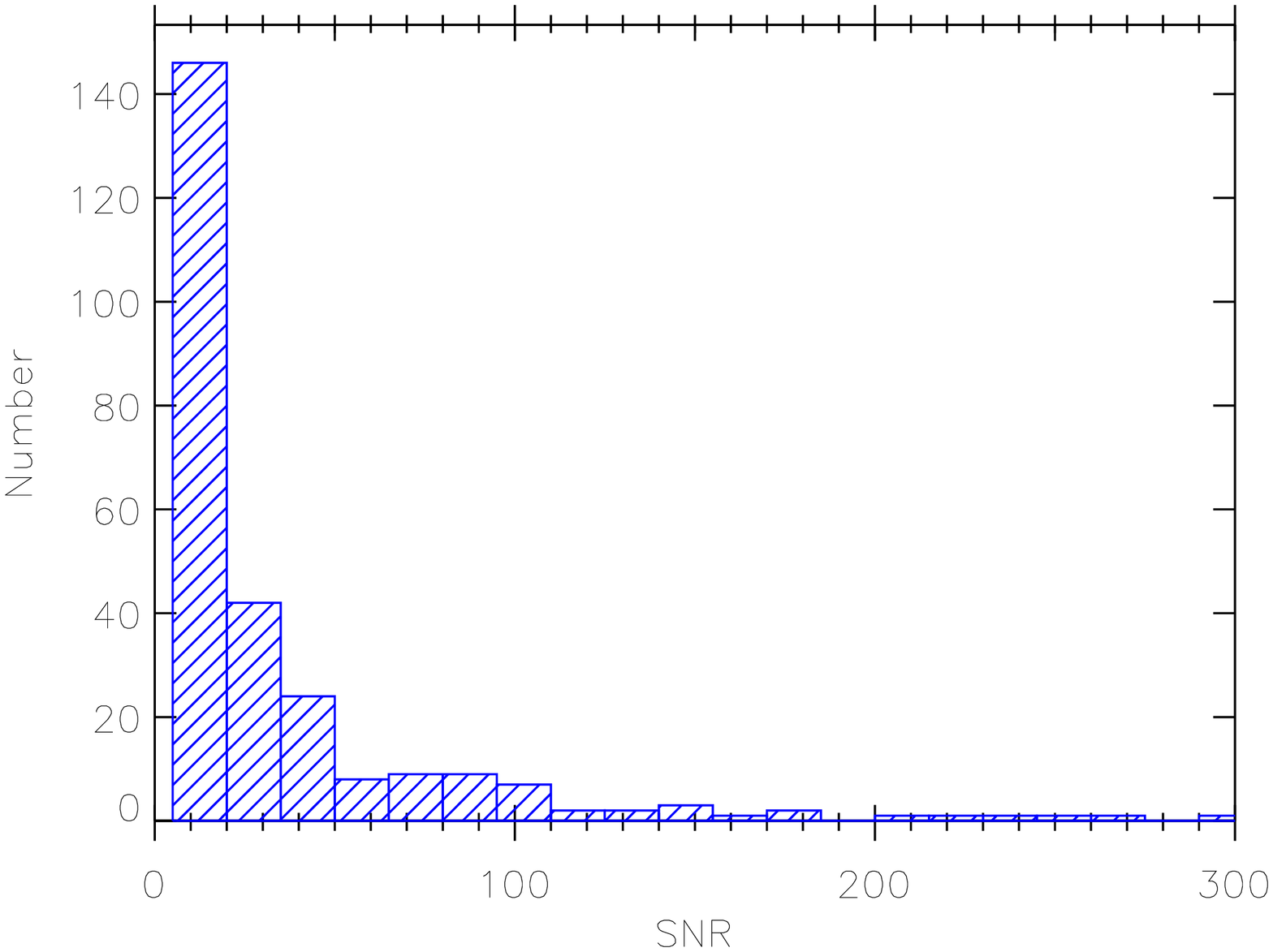}  
  \end{minipage}
  \caption{Distributions of emission redshift (\emph{left}) and SNR
    (\emph{right}) for our full sample of 262 blazars (including the​ ​
    71 blazars which are subsequently excluded from the \mgii
    strong systems analysis; see Section~\ref{subsec:ew_redshiftpath}. For 58
    blazars without exact $z_{em}$, we have used their lower limit on
    $z_{em}$, derived based on the most redshifted absorption
    system seen in their spectra).}
  \label{fig:zemi_snr_dist}
\end{figure*}

The $z_{em}$ and SNR distributions for our sample of 262 blazars
(including 42 from the BBM sample) are shown in
Fig.~\ref{fig:zemi_snr_dist}. However, as described in
Section~\ref{subsec:ew_redshiftpath}, among the  220 non-BBM blazars 
only 149 blazars were found to contribute to robust redshift path for 
the case of strong absorbers and only 58 of them have contributed to the 
robust redshift path for weak absorbers, as well. Thus, for the purpose 
of the present $dN/dz$ analysis, we are led to a final sample of 191 blazars 
(149 ours plus 42 BBM blazars) for the strong absorber case. Only 100 out of 
them also contribute to the $dN/dz$ analysis for weak absorbers. 
The sample of 191 blazars is listed in
Table~\ref{table:full_sample}.
\begin{table}
\caption{Basic properties of our sample of 191 blazars.}
{\scriptsize
  \begin{tabular}{@{}lll lr @{}}
    \hline \hline
Target            & $z_{em}$    & Data archive & $\Delta$z$_{strong}^{a}$ & SNR  \\
\hline \\
J001937$+$202146  &  0.858    &  KECK  &  0.417  &  6.4    \\
J003514$+$151504  &  0.443    &  SDSS  &  0.123  &  55.0    \\
J003808$+$001336  &  0.740    &  FORS/ESO  &  0.439  &  91.8    \\
J004054$-$091525  &  5.030    &  FORS/ESO  &  0.891  &  66.2    \\
---               &   ---     &   ---  & ---     &  -----      \\
\hline
\multicolumn{5}{l}{{ Note.} The entire table is available  in on-line version. Only a}\\
 \multicolumn{5}{l}{ portion of this table is shown here, to display its form and content.}\\
\multicolumn{5}{l}{$^{a}$The redshift path contribution to the strong \mgii system}\\
\multicolumn{5}{l}{(W$_r (2796) \geq 1.0$~\AA) analysis.}\\
\label{table:full_sample}
\end{tabular}
}
\end{table}
\section{Analysis}
\subsection{Data Reduction}
\label{subsec:data_reduction}
Data reduction for the FORS spectra available for our 53 blazars,
including the 42 blazars from BBM, was performed using the ESO-FORS
pipeline (version 5.1.4), by executing it using the
ESOREX \footnote{http://www.eso.org/sci/software/cpl/esorex.html}
algorithm.  The pipeline performs a precise background subtraction on
science frames, does master flat-fielding, rejects cosmic-ray impacts
by employing an optimal extraction technique and then applies
calibrations for wavelength and flux. The Keck/LRIS data were reduced
using a publicly available Lris automated reduction
Pipeline\footnote{http://www.astro.caltech.edu/$\sim$dperley/programs/lpipe.html}
(lpipe) written in {\sc idl}. The UVES and SDSS spectra (total 197 blazars)
were already available in the reduced form. Finally, the spectra taken
with the SAO 6-m telescope for 3 of our blazars (viz, J145127$+$635426,
J165248$+$363212 and J182406$+$565100) were reduced using the standard
IRAF tasks. For post-processing of each one-dimensional spectrum,
which involves steps such as air-to-vacuum wavelength conversion,
heliocentric correction, combining individual exposures for SNR
enhancement and continuum fitting to determine the normalized
spectrum, we have followed the procedure described in
\citet{Chand2004A&A...417..853C}.
\subsection{Computation of the EW detection limit and the corresponding useful redshift path}
\label{subsec:ew_redshiftpath}
For identifying absorption features in a given spectrum, proper
evaluation of noise plays a crucial role as it defines the detection
limit for the features. Therefore, to generate the distribution of rms
noise along the spectrum we used the matched-filtering technique employed
by \citet{Zhu2013ApJ...770..130Z}, which involves the following main
steps:\\ (i) Subtracting unity from each point/pixel of the normalized
spectrum (thus, the mean level of the spectrum becomes zero).\\ (ii)
Using this residual spectrum, generate its amplitude version, by 
plotting only the magnitude of the signal at each spectral pixel (i.e.,
setting the negative sign to positive).\\ (iii) This `noise amplitude
spectrum' is then subjected to a top-hat smoothing over the
`effective spectral resolution', which is taken to be the quadratic
sum of the instrumental resolution and the typical \mgii absorption
line width.
For instance, the typical FWHM of the QSOs absorption line convolved
with the SDSS instrumental resolution lies in the range $\sim$ 100-400 kms$^{-1}$
($\sim$ 2-6 pixels). Thus, for the SDSS instrumental resolution of 120 kms$^{-1}$, 
we have taken an ‘effective spectral resolution’ of about 4 pixels
\citep[i.e., $\sim$
  276 kms$^{-1}$, e.g., see][]{Zhu2013ApJ...770..130Z}.\\ (iv) This
`smoothed noise amplitude spectrum' was then sub-divided into a
sequence of 100~\AA~ wide segments. Within each segment all points
deviating by more than one $\sigma$ were clipped (including any spectral
lines) and substituted with interpolated values. {In this smoothed
noise amplitude spectrum the amplitude at a given spectral
(i.e., wavelength) pixel `$i$' is the representative noise, $n_i$, for that
pixel.}\\ (v) For each spectral pixel, we then set the 3$\sigma$  
detection threshold of a spectral feature. To do this, we model
the feature as a Gaussian having a FWHM equal to the afore-mentioned
`effective spectral resolution' and then subject it to
the same `top-hat' smoothing as mentioned in (iii) above, and
finally, we optimize its amplitude to equal 3$n_i$. Equivalent width of
the Gaussian satisfying this criterion thus becomes the limiting
equivalent width ($W_{i,det}$) of the \mgii absorption line that would be
accepted as a significant (3$\sigma$) detection at that particular pixel
in the spectrum. 
Only provided the pre-set restframe threshold value ($W_{th}$), 
which is 0.3~\AA~ for weak and 1.0~\AA~ for strong absorption systems, 
respectively, exceeds the computed  $W_{i,det}$ for that pixel, 
would that spectral pixel be accepted as contributing
to `useful' redshift path, not otherwise. Note that this procedure is
very similar to that adopted in \citet{2017arXiv170105624M}.\par As a result,
for our blazar sample, the net useful redshift path at a given redshift $z_{i}$
(so that the \mgii absorption line falls in the $i^{th}$ spectral
pixel) for detection of the \mgii doublet above a designated
rest-frame equivalent width threshold, $W_{th}$ would be:
\begin{eqnarray}
\begin{aligned}
  g\left(W_{th},z_{i} \right) & = \sum_{j=1}^{N_{blazar}}\ H(z_i - z_{j,
    min})\times H(z_{j, max} - z_i )\\ & \times H(W_{th} -{W_{j,det}(z_{i})})
  \\
\end{aligned}
\label{eq:goz1}
\end{eqnarray}

where $H$ is the Heaviside step function, and the summation is taken
over all the blazar spectra in our sample, $z_{j,min}$ and
$z_{j,max}$ are, respectively, the minimum and maximum expected
absorption redshift limits which were used in the search for the \mgii
doublet for $j^{th}$ quasar (see Section~\ref{subsec:mgii_iden_mgiidndz}).  
$W_{th}$ is the threshold rest-frame EW of the \mgii absorption line which we have set 
at 1~\AA~ and 0.3~\AA, for strong and weak absorption 
systems, respectively. $W_{j,det}(z_{i})$ is the computed rest-frame EW 
detection limit at the i$^{th}$ pixel in the spectrum of the $j^{th}$ 
quasar, as discussed above. 
In the parent sample of 262 blazar, a non-zero $g\left(W_{th},z_{i} \right)$ 
was found for only 191 blazars for the strong absorber case, and 100 out of
them also contributed a non-zero $g\left(W_{th},z_{i} \right)$ for the case of 
weak absorption system, as well. Hence only these two subsets have been used in 
the present $dN/dz$ analysis.
\begin{figure}
  \includegraphics[width=0.3\textwidth,height=0.35\textheight,angle=90]{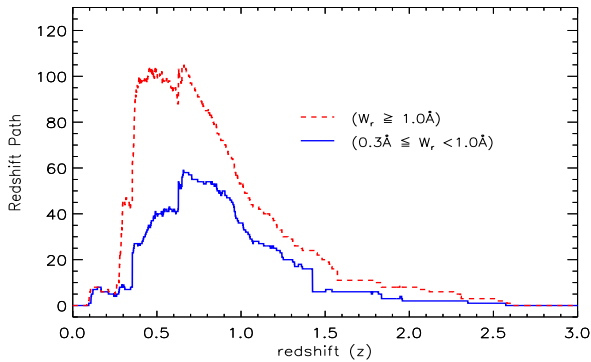}
 \caption{The distributions of redshift path density for the intervening 
   Mg\,{\sc ii} systems towards blazars, for the strong 	
  (W$_r (2796) \geq 1.0$~\AA: red dashed curve) and the weak absorption systems	
  ($0.3 \leq $ W$_r (2796) < 1.0$~\AA: blue solid curve).}	
         \label{fig:redshift_path_density}	
\end{figure}
\subsection{\mgii absorption-line identification}
\label{subsec:mgii_iden_mgiidndz}
\begin{table}
\caption{The detected 43 \mgii absorption systems\\ and their rest-frame 
equivalent widths, W$_r$(\mgii $\lambda 2796$~\AA),\\ as identified in 34 blazars 
belonging to our sample of 191 blazars.}
{\scriptsize
\begin{tabular}{@{}ccccl @{}}
\hline \hline \multicolumn{1}{c}{Blazar} 
& $z_{em}$
& $z_{abs} $
& W$_{r}$  
& Associated absorptions  \\
\hline \\

J001937$+$202146  &  0.858  &  0.69616  &  1.0969 & \feii\\
J010009$-$333730  &  0.875  &  0.67996  &  0.5110 & \feii\\
J014125$-$092843  &  0.730  &  0.50042  &  0.3641 & \feii\\
J021748$+$014449  &  1.715  &  1.34428  &  2.0130 & \mgi, \feii, \znii, \mnii\\
J023838$+$163659  &  0.940  &  0.52379  &  2.3725 &  \mgi, \feii, \mnii\\   
---               &   ---   &   ---     & ---     &  ---\\
\hline
\multicolumn{5}{l}{{ Note.} The entire table is available in the on-line version.}\\
 \multicolumn{5}{l}{ Only a portion of the table is shown here to display its form}\\
\multicolumn{5}{l}{ and content.}\\ 
\label{table:mgii_absorbers}
\end{tabular}
}
\end{table}
Normally, the wavelength coverage may differ from spectrum to spectrum,
as it depends on the spectrograph's specifications and the instrumental 
settings used for the observations. Additionally, we place two constraints 
on the redshift path over which the search for the \mgii doublet was made.
Firstly, the search was restricted to within the range $(1+z_{em}) \times 1216$\AA~ 
$< \lambda < (1+z_{em}) \times 2803$\AA. The lower limit is meant to avoid 
the Ly$\alpha$ forest, while the upper limit is dictated by the fact that 
any \mgii absorber with $\lambda \ge (1+z_{em}) \times 2803$\AA~ has a strong 
likelihood of being an associated system falling into the background AGN. 
We performed the search for the \mgii doublet at $z_{em}$ following the 
steps enumerated in section 3 of \citet{Chand2012ApJ...754...38C}.  
Accordingly, a line profile matching technique was used, such that we first 
plotted the normalized spectrum of a given blazar and then overplotted the 
same spectrum by shifting the wavelength axis by a factor of
\mgii$\lambda2796.3543$/\mgii$\lambda2803.5315$ (i.e. 0.997). Then,
about 50~\AA~ wide spectral segments were manually examined. The
location of perfect overlap between the absorption lines in the
shifted and the original (unshifted) spectra were marked as a detected 
\mgii absorption system.
As a further corroboration, we looked for the corresponding metal lines
(e.g. {\feii}, {\civ}, {\siiv}, etc.) in the spectrum. If the 
redshift/velocity of a \mgii absorption doublet was found to be consistent 
with that of the peak estimated for the metal line(s), this was taken as 
a further
confirmation of the previously identified \mgii absorber. Note that the
systems having a velocity offset within 5000 kms$^{-1}$ of $z_{em}$
of the blazar were classified as associated systems, following the
standard practice.  For each detected \mgii absorption system, we also
performed visually a quality check on the fit to the underlying
continuum. If deemed desirable we carried out a local continuum
fitting and this improved fit was then used to obtain a better
estimate of \ew(\mgii). Detailed information on the 43 \mgii
absorption systems thus identified in the spectra of 34 blazars, out of the total
present sample of $191$ blazars, is provided in
Table~\ref{table:mgii_absorbers}.  \par
\subsection{Computation of $dN/dz$ }
The incidence rate of Mg~{\sc ii} absorbers is defined as $\frac{dN}{dz}
= N_{obs}/\Delta z$; where $N_{obs}$ is the total number of the Mg~{\sc
ii} absorbers detected within the entire {\it useful} redshift path ($\Delta z$), 
defined as:
\begin{equation}
  \Delta z = \int_0^\infty \sum_{j=1}\ g_{j}(W_{min},z_i) dz_{i}
  \label{eq:goz}
\end{equation}
where g$_{j}(W_{min},z_i) = 1$ if W$_{th}$ ( 0.3~\AA~ for weak systems and
1~\AA~ for strong systems) is more than the (3$\sigma$) detection
threshold W$_{j,det}(z_j)$ estimated for the $i^{th}$ spectral pixel
(see Eq.~\ref{eq:goz1}), otherwise $g(W_{min},z_i) = 0$. 
The values of redshift path density for $i^{th}$ redshift pixel, i.e., 
$g(W_{min},z_i) $, were thus computed using the total 191 blazar sightlines 
for strong absorption systems and the 100 blazar sightlines for weak absorption
systems (Fig.~\ref{fig:redshift_path_density}).\par
Table~\ref{table:full_sample} lists the values of $\Delta z$
calculated for the individual sightline in our entire blazar sample and in its various sub-sets. The 
errors in the computed $dN/dz$ values were estimated
assuming the Poisson small number statistics for $N_{obs} < 50$, within a
limit of 1$\sigma$ confidence level of a Gaussian distribution, using
the tabulation given by \citet{Gehrels1986ApJ...303..336G}.
\section{Results}
\subsection{Incidence of \mgii absorbers in the present enlarged sample}
\label{section:enlarge_sample_analysis}
\begin{table*}
 {\scriptsize
\caption{Incidence of \mgii absorbers for the present sample and its various subsets.}
\centering
\begin{tabular}{llllclllc}
\toprule 
    Sample type & \multicolumn{4}{c}{Weak absorption systems} & \multicolumn{4}
	{c}{Strong absorption systems}\\\\    
    & $N_{obs}$ & $\Delta$z & $\frac{N_{obs}}{\Delta z} \equiv dN/dz$ & $\left(\frac{(N_{obs}/\Delta z)}{(dN/dz)_{qso}} \right)^{\alpha}$ 
    & $N_{obs}$ & $\Delta$z & $\frac{N_{obs}}{\Delta z}$ & $\left(\frac{(N_{obs}/\Delta z)}{(dN/dz)_{qso}} \right)^{\beta}$ \\\\
  \midrule
Sample I    $^{a}$     &  23&  44.21& $  0.52_{  0.11}^{  0.13}  $& $ 1.21_{  0.25}^{  0.31}  $&20&  86.08&$  0.23_{  0.05}^{  0.06} $&$   1.26_{  0.28}^{  0.35}  $ \\\\
Sample II   $^{b}$     &  20&  40.06& $  0.50_{  0.11}^{  0.14}  $& $ 1.16_{  0.26}^{  0.32}  $&16&  81.41&$  0.20_{  0.05}^{  0.06} $&$   1.07_{  0.26}^{  0.34}  $ \\\\
Sample III  $^{c}$     &  17&  32.69& $  0.52_{  0.12}^{  0.16}  $& $ 1.20_{  0.29}^{  0.37}  $&13&  60.10&$  0.22_{  0.06}^{  0.08} $&$   1.11_{  0.30}^{  0.40}  $ \\\\
Sample IV   $^{d}$     &  6&   19.05& $  0.31_{  0.12}^{  0.19}  $& $ 0.74_{  0.29}^{  0.44}  $&8&   58.98&$  0.14_{  0.05}^{  0.07} $&$   0.80_{  0.28}^{  0.39}  $ \\\\
Sample V    $^{e}$     &  3&   11.69& $  0.26_{  0.14}^{  0.25}  $& $ 0.57_{  0.31}^{  0.56}  $&5&   37.67&$  0.13_{  0.06}^{  0.09} $&$   0.72_{  0.31}^{  0.49}  $ \\\\
Sample VI   $^{f}$     &  4&   15.48& $  0.26_{  0.12}^{  0.20}  $& $ 0.59_{  0.28}^{  0.47}  $&7&   52.21&$  0.13_{  0.05}^{  0.07} $&$   0.76_{  0.28}^{  0.41}  $ \\\\
Sample VII  $^{g}$     &  2&   9.29 & $  0.22_{  0.14}^{  0.28}  $& $ 0.46_{  0.29}^{  0.60}  $&5&   32.70&$  0.15_{  0.07}^{  0.10} $&$   0.76_{  0.33}^{  0.51}  $  \\\\
Sample VIII $^{h}$     &  16&  24.58& $  0.65_{  0.16}^{  0.21}  $& $ 1.52_{  0.38}^{  0.48}  $&9&   29.20&$  0.31_{  0.10}^{  0.14} $&$   1.61_{  0.53}^{  0.73}  $ \\\\
Sample IX   $^{i}$     &  15&  23.41& $  0.64_{  0.16}^{  0.21}  $& $ 1.50_{  0.38}^{  0.50}  $&8&   27.41&$  0.29_{  0.10}^{  0.14} $&$   1.52_{  0.52}^{  0.75}  $ \\\\
\hline

\multicolumn{9}{l}{$^{\alpha}$ $dN/dz $ for QSOs is calculated at the mean value of the redshift path of the blazars, using equation 2 of BBM.} \\
\multicolumn{9}{l}{$^{\beta}$ $dN/dz$ for QSOs is calculated at the mean value of the redshift path of the blazars, using equation 6 of BBM. }\\\\
\multicolumn{9}{l}{ $^{a}$ Full sample of 191 blazars (also included are the sources with only a lower limit available for ~$z_{em}$).}\\  
\multicolumn{9}{l}{ $^{b}$ 184 BL Lacs, i.e., Sample I (191 blazars) $-$ 7 (BBM FSRQs).}\\ 
\multicolumn{9}{l}{ $^{c}$ 133 BL Lacs, i.e., Sample II ( 184 BL Lacs) $-$ 51 (sources with only a lower limit on $z_{em}$).}\\  
\multicolumn{9}{l}{ $^{d}$ 149 Non-BBM BL Lacs, i.e., Sample I (191 blazars) $-$ 42 (BBM blazars).}\\ 
\multicolumn{9}{l}{ $^{e}$ 98 Non-BBM BL Lacs, i.e., Sample IV (149 Non-BBM BL Lacs) $-$ 51 (Non-BBM BL Lacs with only lower limit on $z_{em}$).} \\  
\multicolumn{9}{l}{ $^{f}$ 126 SDSS BL Lacs, i.e., Sample II (184 BL Lacs) $-$ 58 (Non-SDSS BL Lacs).}\\ 
\multicolumn{9}{l}{ $^{g}$ 79 SDSS BL Lacs, i.e., Sample VI (126 SDSS BL Lacs) $-$ 47 (SDSS BL Lacs with only lower limit on $z_{em}$).}\\  
\multicolumn{9}{l}{ $^{h}$ 58 Non-SDSS BL Lacs, i.e., Sample II (184 BL Lacs) $-$ 126 (SDSS BL Lacs).} \\  
\multicolumn{9}{l}{ $^{i}$54 Non-SDSS BL Lacs, i.e., Sample VIII (58 Non-SDSS BL Lacs) $-$ 4 (Non-SDSS BL Lacs with only lower limit on $z_{em}$).}\\ 

  \end{tabular}
  \label{Table:all_dndz_result}
}
\end{table*}

\begin{figure*}
  \hspace{-1.1in}
  \begin{minipage}[]{0.4\textwidth}
  \includegraphics[width=0.8\textwidth,height=0.4\textheight,angle=90]{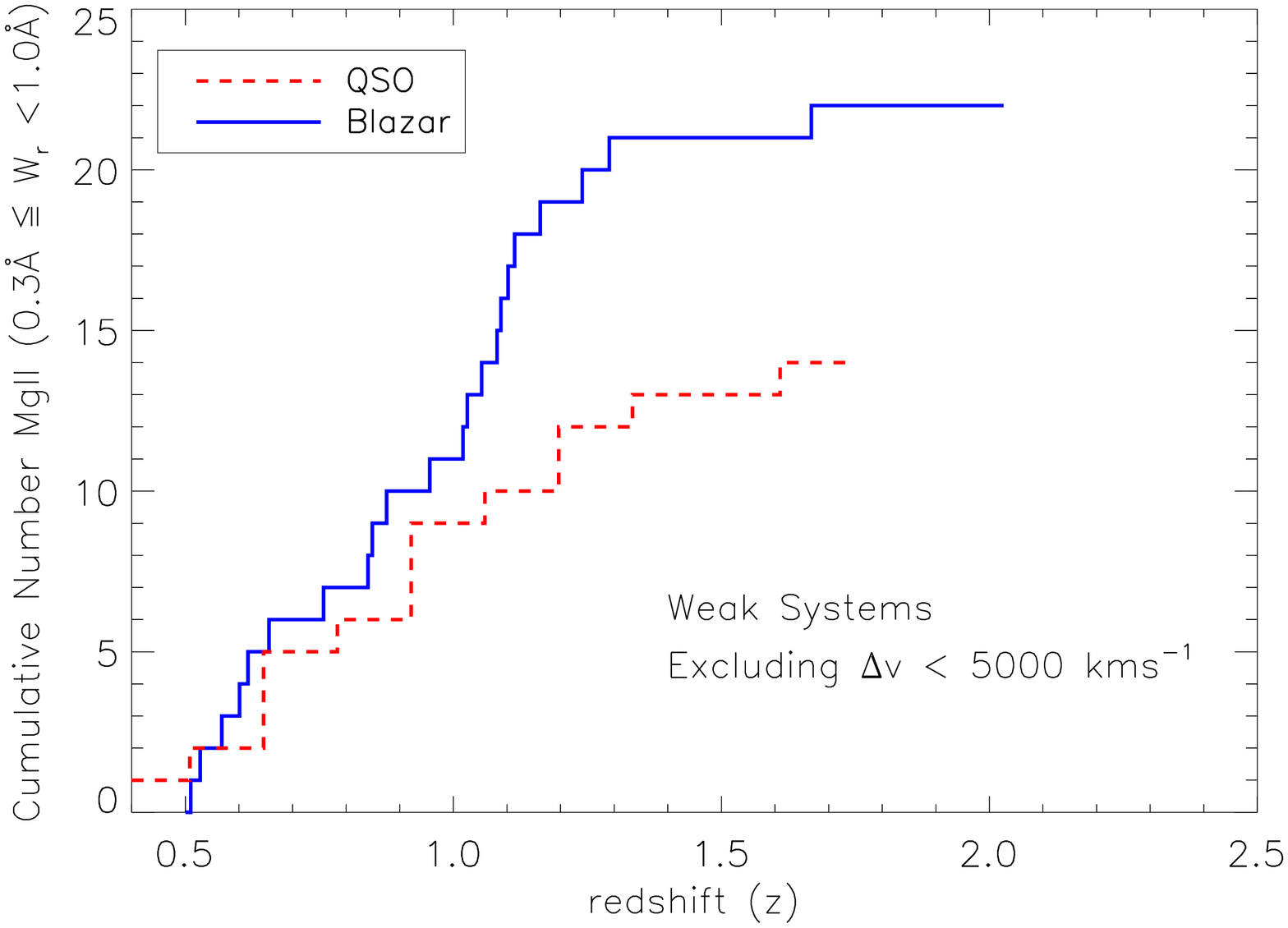}    
  \end{minipage}
  \hspace{0.8in}
  \begin{minipage}[]{0.4\textwidth}
  \includegraphics[width=0.8\textwidth,height=0.4\textheight,angle=90]{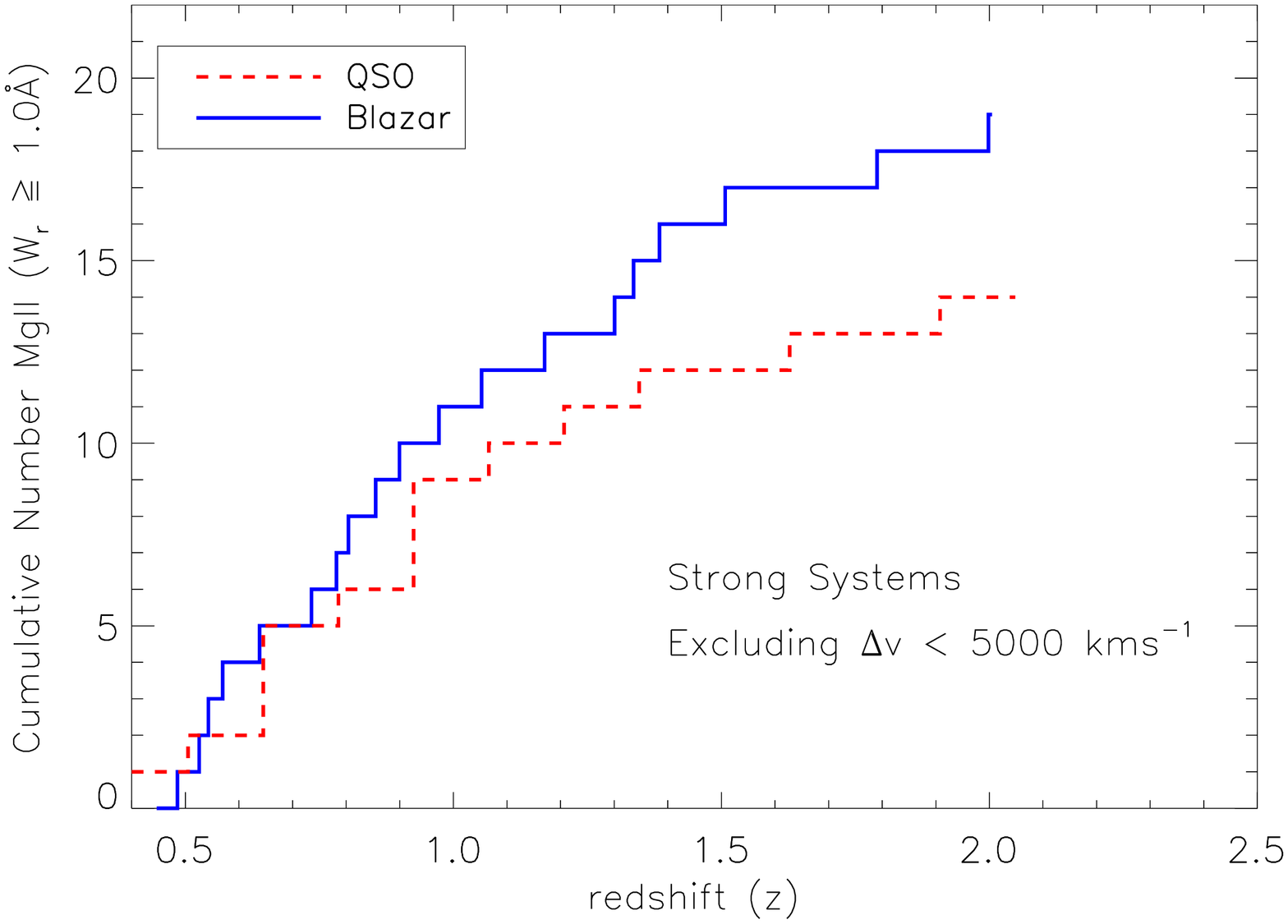}  
  \end{minipage}\\
  \vspace{-0.2in}
  \hspace{-1.1in}
  \begin{minipage}[]{0.4\textwidth}
  \includegraphics[width=0.8\textwidth,height=0.4\textheight,angle=90]{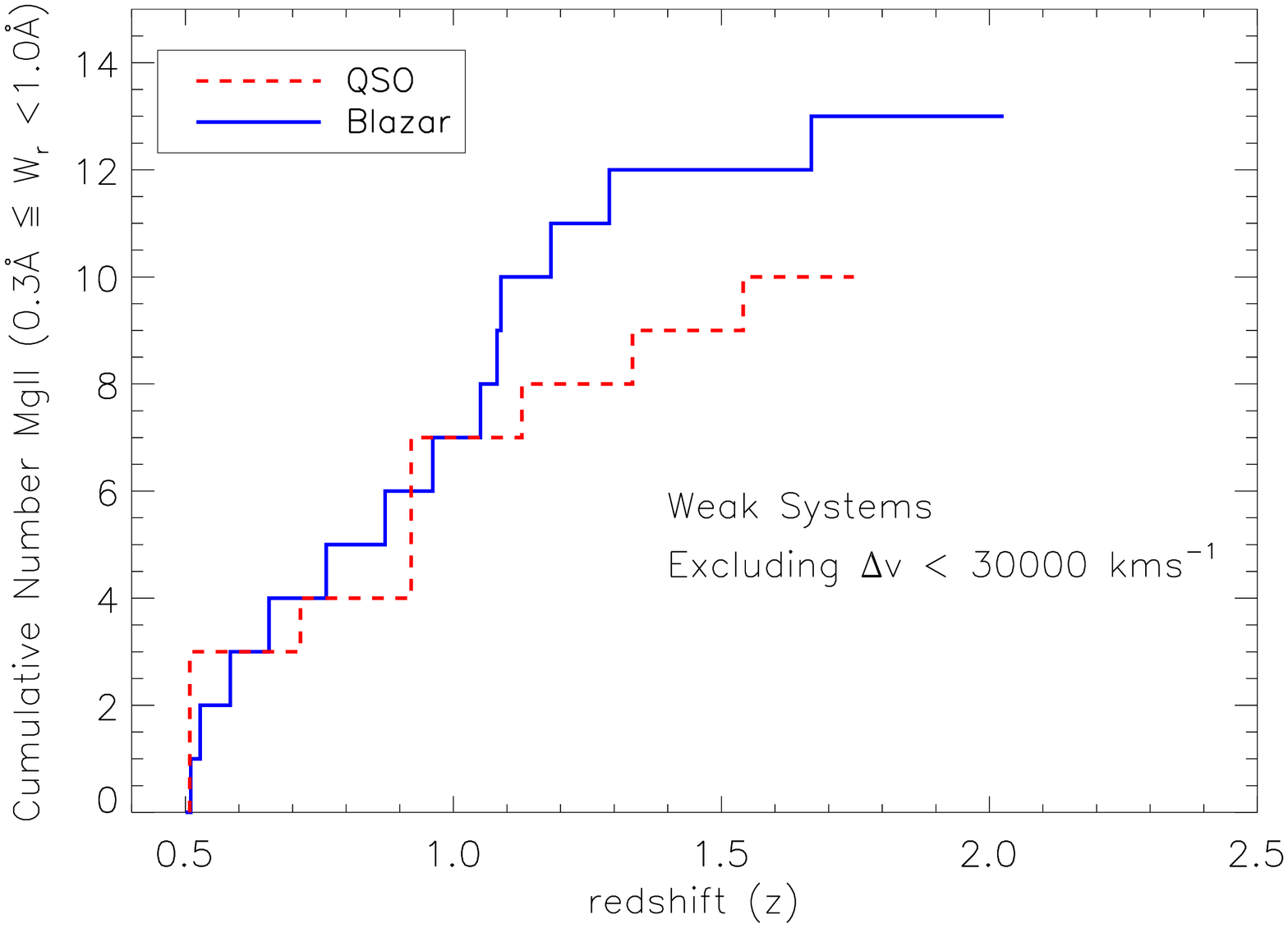}    
  \end{minipage}
  \hspace{0.8in}
  \begin{minipage}[]{0.4\textwidth}
  \includegraphics[width=0.8\textwidth,height=0.4\textheight,angle=90]{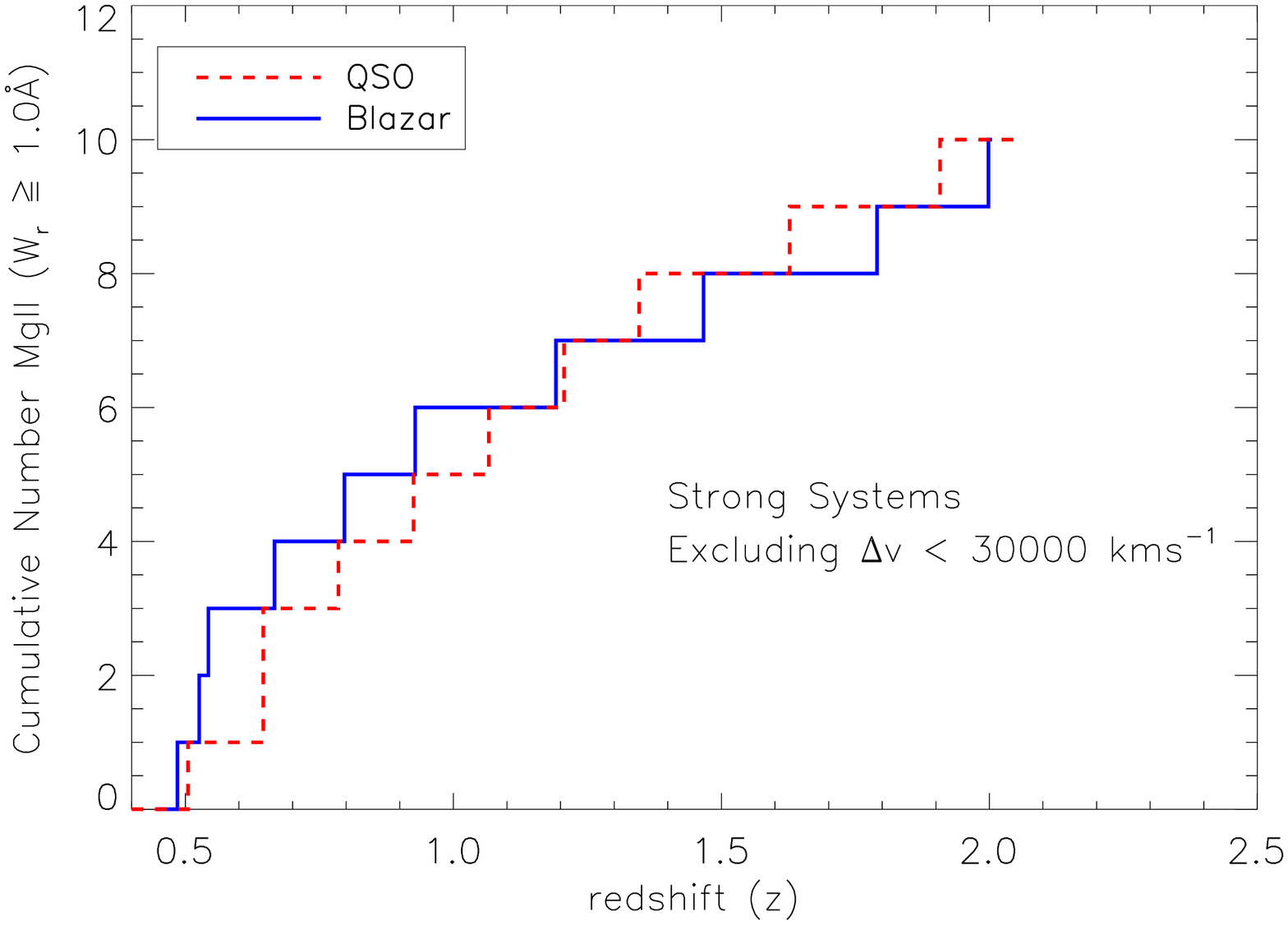}  
  \end{minipage}\\
  \vspace{-0.2in}
  \hspace{-1.1in}
  \begin{minipage}[]{0.4\textwidth}
  \includegraphics[width=0.8\textwidth,height=0.4\textheight,angle=90]{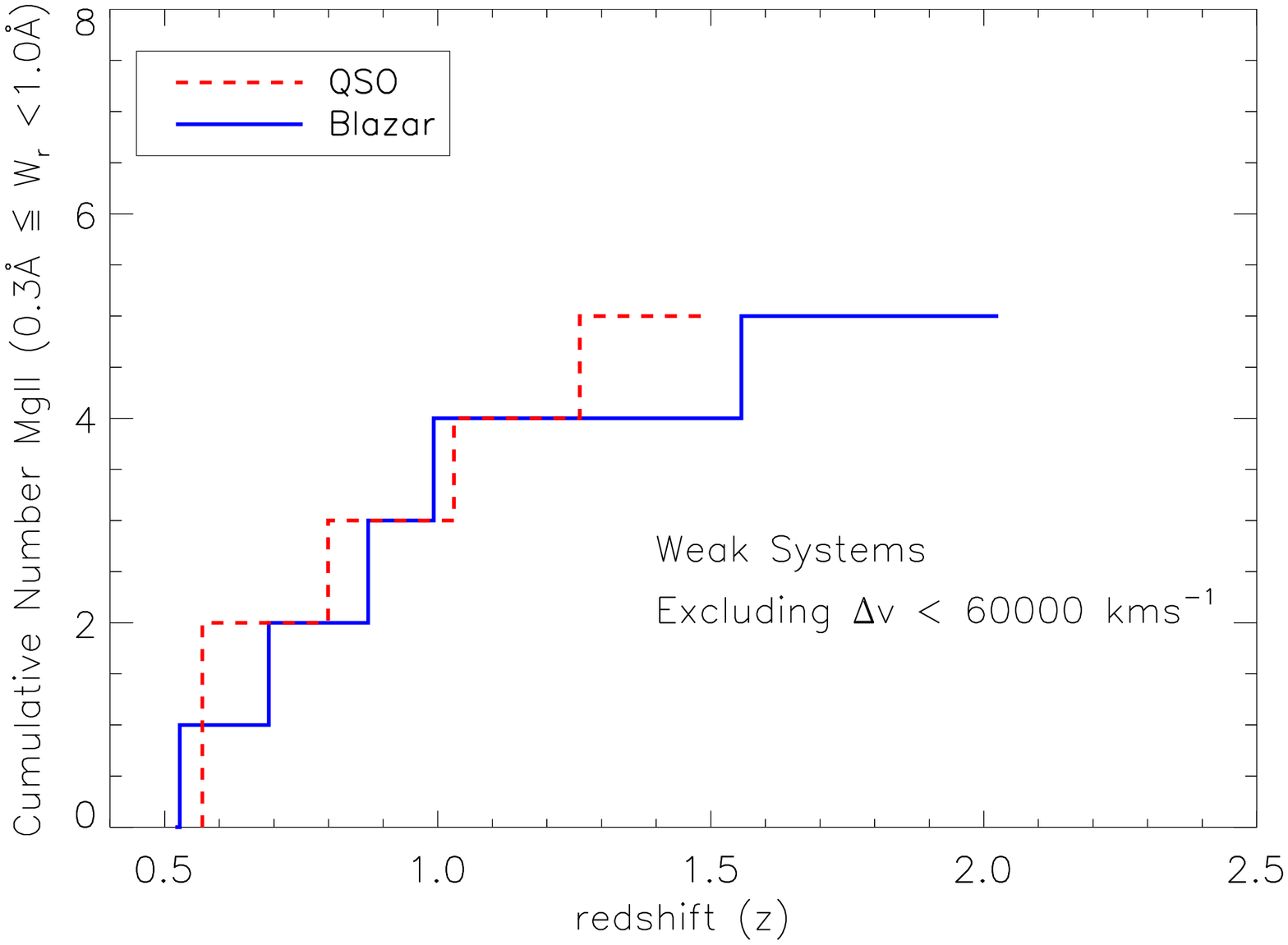}    
  \end{minipage}
  \hspace{0.8in}
  \begin{minipage}[]{0.4\textwidth}
  \includegraphics[width=0.8\textwidth,height=0.4\textheight,angle=90]{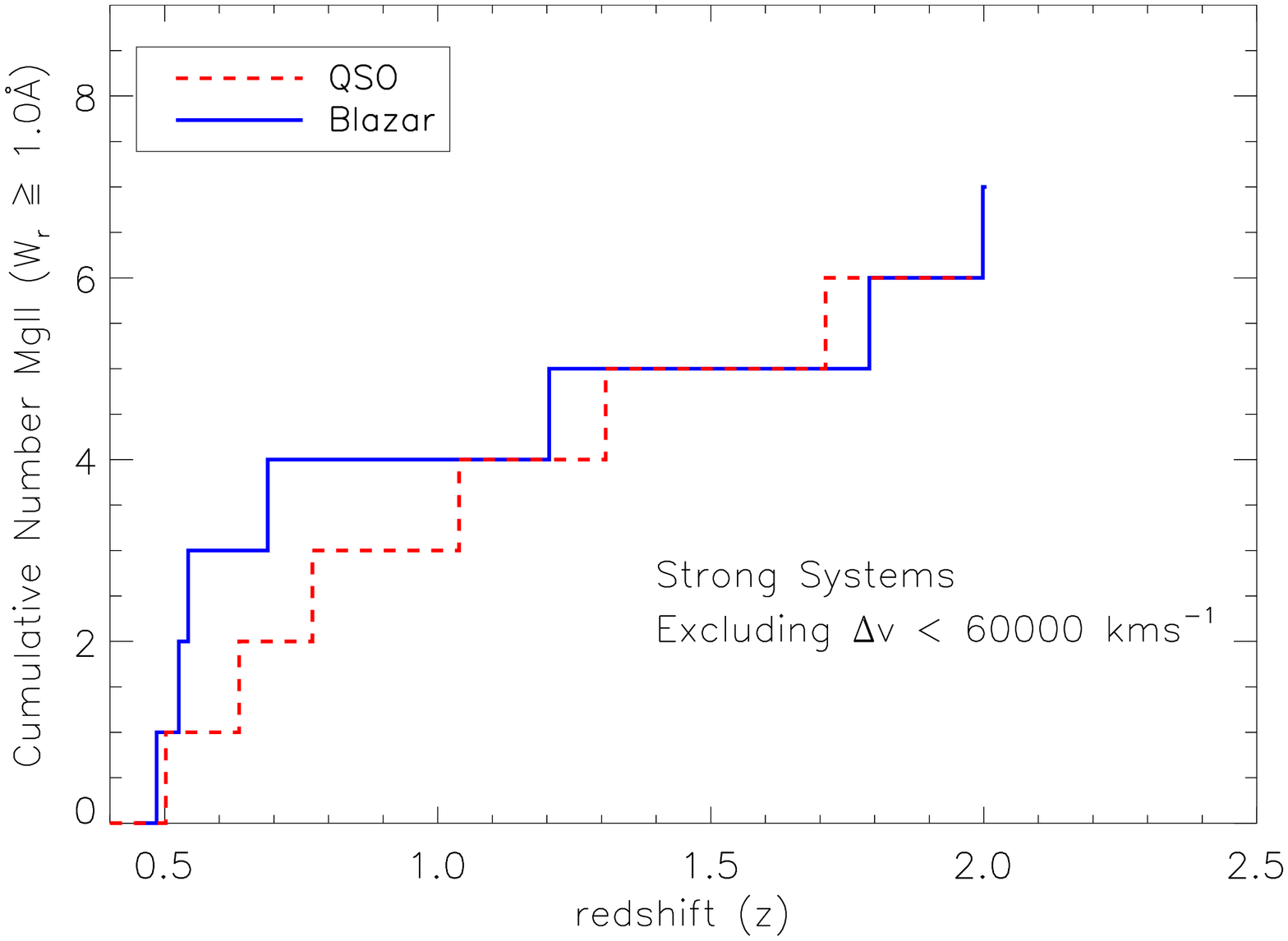}  
  \end{minipage}
  \caption{Cumulative number of weak (\emph{left}) and strong
    (\emph{right}) intervening Mg\,{\sc ii} absorption systems detected towards
    blazars (blue solid line) and QSOs (red dashed line), after
    excluding the systems with offset velocity ($\Delta v$) $< 5000$
    kms$^{-1}$ (\emph{top panel}), $\Delta v < 30000$ kms$^{-1}$
    (\emph{middle panel}) and $\Delta v < 60000$ kms$^{-1}$
    (\emph{bottom panel}).  For QSOs the estimates for weak \mgii
    systems are taken from \citet{Nestor2005ApJ...628..637N} (see
    text), and for strong systems these are adopted from
    \citet{Prochter2006ApJ...648L..93P}}.
         \label{fig:cummulative_plot}
\end{figure*}

\begin{table} 
\begin{minipage}[10]{130mm}
\caption{\small Incidence of strong Mg~{\sc ii} absorption
  systems\newline towards the low SNR (SNR$<$15) and high SNR
  ($>$15)\newline spectra in our sample of 191 blazars.  The last
  column shows \newline the same relative to normal QSOs.}
{\scriptsize
\begin{tabular}{@{}rccc ccccc @{}}
\hline \hline
\multicolumn{1}{c}{Sample type}
& \multicolumn{1}{c}{$N_{obs}$}
& $\Delta$z
&$\frac{N_{obs}}{\Delta z}$
&$\left(\frac{(N_{obs}/\Delta z)}{(dN/dz)_{qso}} \right)^{\alpha}$\\
\hline \\
Low SNR   &  7&   28.74&$   0.24_{  0.09}^{  0.13}  $&$   1.19_{  0.44}^{  0.64}  $  \\\\          
High SNR  &  13&  57.33&$   0.23_{  0.06}^{  0.08}  $&$   1.29_{  0.35}^{  0.47}  $  \\\\          
\hline
\multicolumn{5}{l}{$^{\alpha}(dN/dz)_{qso}$ for the strong systems (W$_r (2796) \geq 1.0$~\AA) is calculated}\\
\multicolumn{5}{l}{(as done in BBM analysis) based on \citet{Prochter2006ApJ...648L..93P}.}\\
\end{tabular}
\label{table:low_high_snr_comp}
}
\end{minipage}
\end{table}
As noted in Section~\ref{sec:intro_mgiidndz}, the BBM result is based
on a sample of just 45 blazars and this has motivated us to build an
enlarged sample. Our sample of 191 blazars provides nearly a factor of 3 
increase in the redshift path (Table~\ref{table:full_sample}). 
Table~\ref{Table:all_dndz_result} summarizes the results for this enlarged 
sample and its various subsets. Note that, as in BBM, the values of $dN/dz$ 
for normal QSOs are calculated at the mean value of the redshift path for
the corresponding blazar subset (column 1 of Table~\ref{Table:all_dndz_result}).
This was done using equation 2 of BBM
for the case of weak absorption systems, and equation 6 of BBM,
for the case of strong absorption systems. From
Table~\ref{Table:all_dndz_result}, no significant excess is evident in
the $dN/dz$ for the blazar sightlines, vis a vis normal QSOs, both for
weak (column 5) and strong (column 9) \mgii absorbers. The same is
apparent from Fig.~\ref{fig:cummulative_plot}, which displays the
cumulative numbers of \mgii absorbers up to different values of
absorption redshift. Although, when only the absorption systems
  having $\Delta v < 5000$ kms$^{-1}$ are deemed as associated systems
  and therefore excluded,  a mild excess may be present for blazar
  sightlines (Fig.~\ref{fig:cummulative_plot} top panel). However, it vanishes
  for the strong systems if one excludes all absorbers having offset
  velocities up to $\Delta v < 30000$ kms$^{-1}$
  (Fig.~\ref{fig:cummulative_plot} middle panel).
 The mild excess vanishes even for weak systems if the absorbers with
 offset velocities up to $\Delta v < 60000$ kms$^{-1}$ are excluded
 (Fig.~\ref{fig:cummulative_plot} bottom panel), suggesting the possibility of extension of
  \mgii intrinsic absorbers up to $\Delta v$ = 0.2c for blazars.
 In any case, our focus here is on the results for strong absorption systems, 
which are statistically more robust since the majority of our spectra (which 
have relatively low SNR) have contributed to the useful redshift path only
for strong systems and not for weak systems.\par
\begin{figure}
  \includegraphics[width=0.3\textwidth,height=0.35\textheight,angle=90]{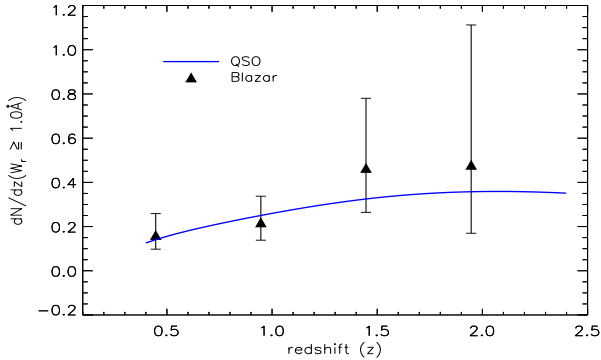}
  \caption{ Number density evolution of strong (W$_{\rm r}(2796) \geq
    1.0$~\AA) \mgii absorption systems (averaged over redshift bins of
    0.5) towards our 191 blazar sightlines (black triangles), and
    the sightlines towards the QSOs in the SDSS (blue solid
    line). The absorption systems with $\Delta v < 5000$ kms$^{-1}$
    have been excluded. The solid line for the SDSS QSOs has been
    computed from the analytical expression given by
    \citet{Zhu2013ApJ...770..130Z} for strong \mgii absorption systems
    towards QSOs. The two distributions show an excellent  agreement with a P$_{null}$ of  $\sim 0.99$
based on the KS and $\chi^{2}$ tests.}
  \label{fig:evol_dndz_zhu}	
\end{figure}

 
In Fig.~\ref{fig:evol_dndz_zhu} we have displayed the redshift
dependence of \dndz for strong ($W_{\rm r}(\lambda 2796) \ge 1.0$~\AA)
\mgii absorption systems detected in our blazar sample and compared it with 
that computed for the sightlines towards normal QSOs, using the analytical 
expression given by \citet{Zhu2013ApJ...770..130Z} for strong \mgii absorption 
systems. To quantify the similarity of these two distributions we have
  applied Kolmogorov-Smirnov (KS) test, which resulted in
  P$_{null}$=0.997; where P$_{null}$ is the null probability that two
  distributions are indistinguishable. Similarly a very good
  statistical agreement is found between the estimates of $dN/dz$ for
  blazars and QSOs, with a $\chi^{2}$-test giving P$_{null}$ = 0.99,
  leading us to infer that the  distributions of $dN/dz$ for blazars
  and QSOs are statistically indistinguishable.
Recall that it is for the strong absorption systems that BBM had
reported a significant excess of $dN/dz$ (compared to QSOs
sightlines), based on the high SNR spectroscopic data available for
their sample of 45 blazars. To pursue this further, we present in the
next section a re-analysis of their data following our data reduction
and analysis procedure.  Since this has yielded results consistent
with the BBM claim, could then the discrepant result we have found
here using a much larger sample of blazars (Table~\ref{table:sample})
have its origin in the substantially lower SNR of the spectra
available for most of the present enlarged sample?  To check this
possibility we divide our blazar sample into (i) a low SNR subset
(spectra having SNR between 5 and 15) and (ii) a high SNR subset (SNR
$\textgreater$ 15).  It is seen from
Table~\ref{table:low_high_snr_comp} that in neither case is a
significant excess of $dN/dz$ (vis a vis normal QSOs) detected for the
strong \mgii absorption systems (the same is found to hold for the
weak absorption systems as well ).  Thus, the discrepant result found
here for the present blazar sample from the BBM sample is unlikely to
be on account of the SNR contrast between the spectral data employed
in the two studies. An alternative possibility is explored in the next
section.

\subsection{Re-analysis of the BBM sample}
\label{subsection:reanalysis_bbm}
\begin{table*}
\caption{Re-analysis of the (high sensitivity) spectra of the 42 southern blazars in the BBM sample.}
{\scriptsize
\begin{tabular}{@{}rccc ccccc @{}}
\hline \hline \multicolumn{1}{c}{Absorber type} &
\multicolumn{1}{c}{W$_{r}(2796)$-range} &
\multicolumn{1}{c}{N$_{re-analysis}$} &
$\Delta$z$_{re-analysis}$  &
\multicolumn{1}{c}{N$_{BBM}$}
& $\Delta z_{BBM}$
& $\left(\frac{(dN/dz)_{re-analysis}}{(dN/dz)_{BBM}} \right) $
\\ \hline \\ Strong$^{\gamma}$ & W$_{r} \ge 1.0 ~$\AA$ $ & 12& 27.10 &
12 & 28.04 & $1.03\pm0.48$ \\\\ Strong$^{\beta}$ & W$_{r} \ge 1.0
~$\AA$ $ & 8& 22.44 & 10 & 23.44 & $0.83\pm0.46$ \\\\ Weak$^{\gamma}$
& $~ 0.3 ~$\AA$ \le $W$_{r}< 1.0$ & 17& 25.16 & 19 & 25.11 &
$0.89\pm0.33$ \\\\ Weak$^{\beta}$ & $~ 0.3 ~$\AA$ \le$ W$_{r}< 1.0$ &
14& 21.01 & 15 & 20.55 & $0.91\pm0.38$ \\\\ \hline
\multicolumn{6}{l}{$^{\gamma}$For the 42 BBM blazars.}\\
\multicolumn{6}{l}{$^{\beta}$ Analysis using 35 BL Lacs in BBM sample.  It is to be noted that out
  of the 42 blazar in BBM sampe,}\\
\multicolumn{6}{l}{32 were designated as BL Lacs. The
  other three included BL Lacs are those which in BBM sample were classified}\\
\multicolumn{6}{l}{ as non-BL Lac (i.e 'opt' class in BBM) viz, J023405$-$301519, J024156$+$004351, J221450$-$293225, but are reported}\\
\multicolumn{6}{l}{as confirmed BL Lacs in VV catalouge and hence are included here in this sample of 35 BL Lacs.}\\
\label{Table:bbm_excess_netzor}
\end{tabular}
}
\end{table*}
As discussed in Section~\ref{sec:intro_mgiidndz}, based on a sample of
45 blazars having high-sensitivity (ESO/FORS) spectra BBM found about
a factor of 2 excess in the number density of \mgii absorbers on the
blazar sightlines, as compared to the sightlines to normal QSOs. Since
the present analysis of a sample of 191 blazars does not show such a
trend, we have carried out a re-analysis of the BBM sample using the
same procedure which we have followed here for our sample. As
mentioned in Section~\ref{sample}, we have limited the re-analysis to
42 out of the 45 BBM blazars, since we could not access the requisite
raw spectral data for the remaining 3 (northern) blazars. For both
weak and strong absorption systems,
Table~\ref{Table:bbm_excess_netzor} compares our results with the BBM
estimates of $N_{obs}$, $\Delta z$ and $dN/dz$.  It is clear that the
BBM estimates are reasonably well reproduced in our analysis; a few
minor discrepancies are noted below.\par For strong systems, there is
a small difference in the redshift path, our value of 27.1 is slightly
lower than the BBM estimate of 28.04. This small difference might be
owing to the difference in the methods of determining \emph{`useful'}
redshift path. We also compared the absorption redshifts and
equivalent widths of the detected \mgii absorption systems and a good
match was found, except in two cases: (i) the system at $z_{abs}$ =
0.5592 towards the blazar J0428$-$3756 was classified as `weak' in BBM
(\ew = 0.93), but `strong' (\ew = 1.03) in our analysis, and (ii) the
$z_{abs}$ = 1.1158 system towards J2031$+$1219 was classified as
`strong' (\ew = 1.29 ) in BBM, but `weak' (\ew = 0.94) in our
analysis. Coming to the weak systems, we detected a total of 17 \mgii
absorbers, whereas BBM reported 19 \mgii absorbers, with the redshift
path being 25.16 in our case, very close to their estimate of
25.11. Two systems, viz. (i) the $z_{abs}$ = 1.1039 system towards
J1419$+$0445 (\ew = 0.52) and (ii) the $z_{abs}$ = 0.6236 system
towards J1956$-$3225 (with \ew= 0.95), could not be included in our
analysis. The former remained undetected and the latter system
corresponds to a wavelength of 4539~\AA~which falls just below the
starting wavelength of 4540~\AA~of the spectrum used in our
analysis.\par
\begin{figure*}
  \hspace{-1.1in}
  \begin{minipage}[]{0.4\textwidth}
  \includegraphics[width=0.8\textwidth,height=0.4\textheight,angle=90]{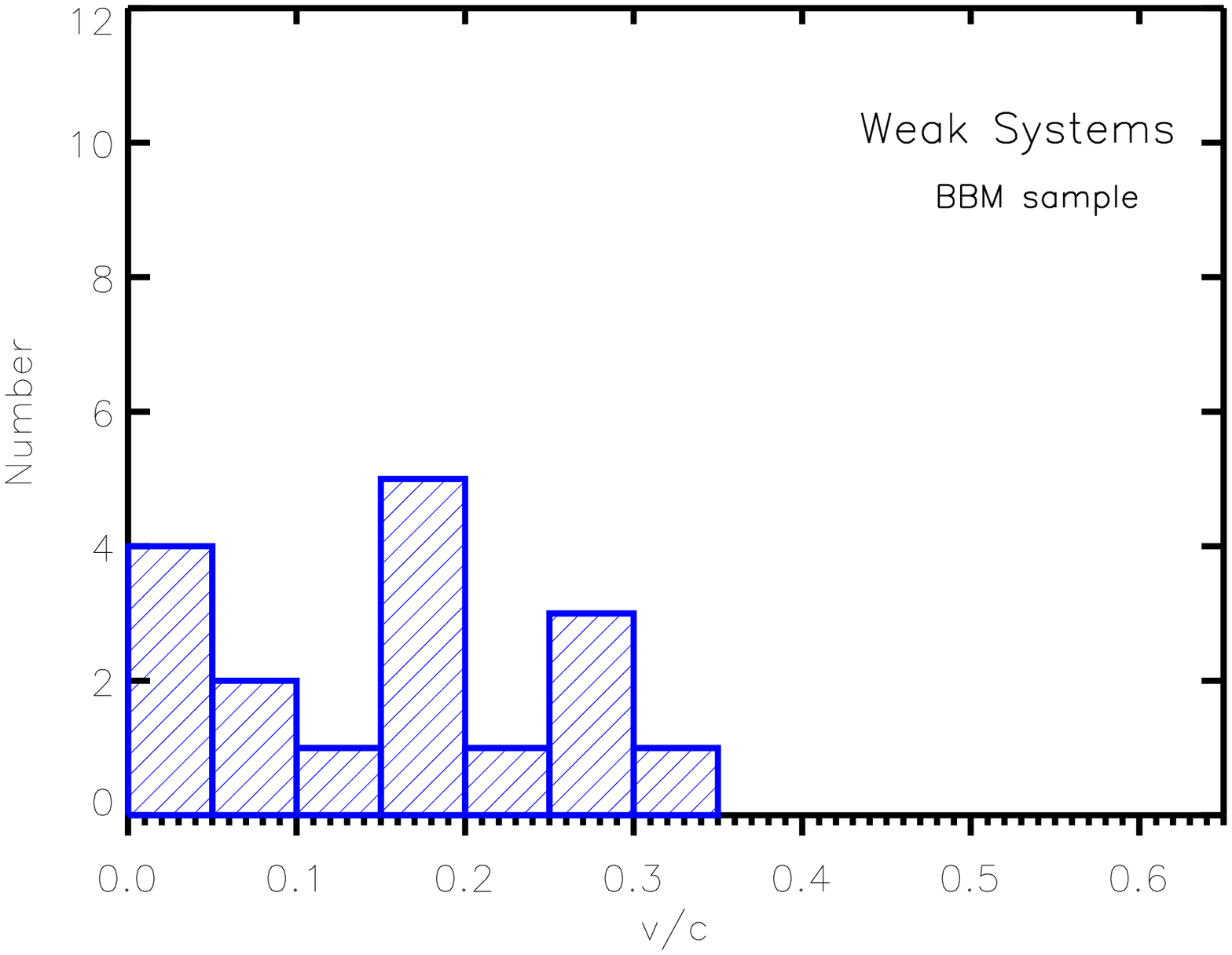}    
  \end{minipage}
 \hspace{0.8in}
  \begin{minipage}[]{0.4\textwidth}
  \includegraphics[width=0.8\textwidth,height=0.4\textheight,angle=90]{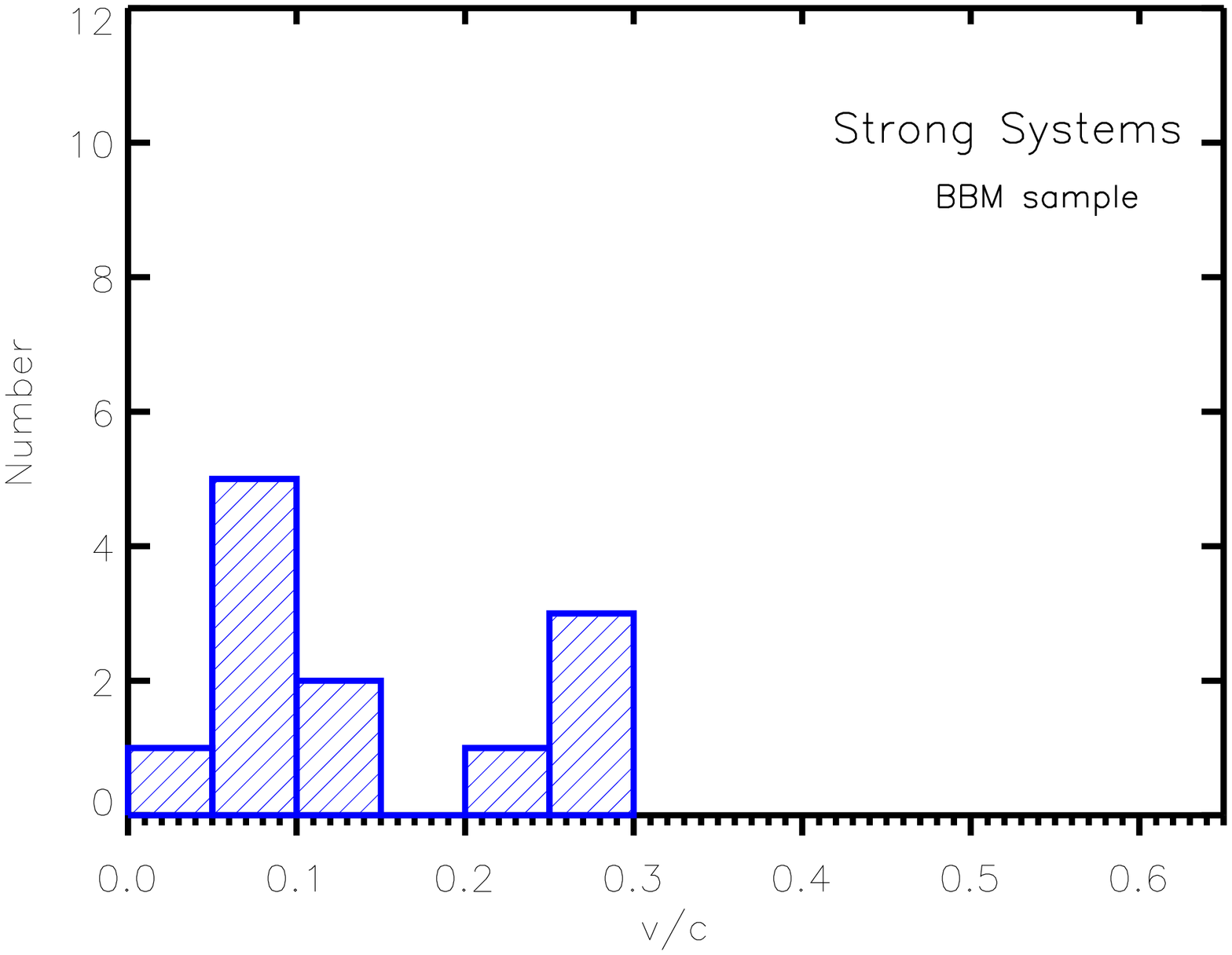}  
  \end{minipage}\\
  \vspace{-0.2in}
  \hspace{-1.1in}
  \begin{minipage}[]{0.4\textwidth}
  \includegraphics[width=0.8\textwidth,height=0.4\textheight,angle=90]{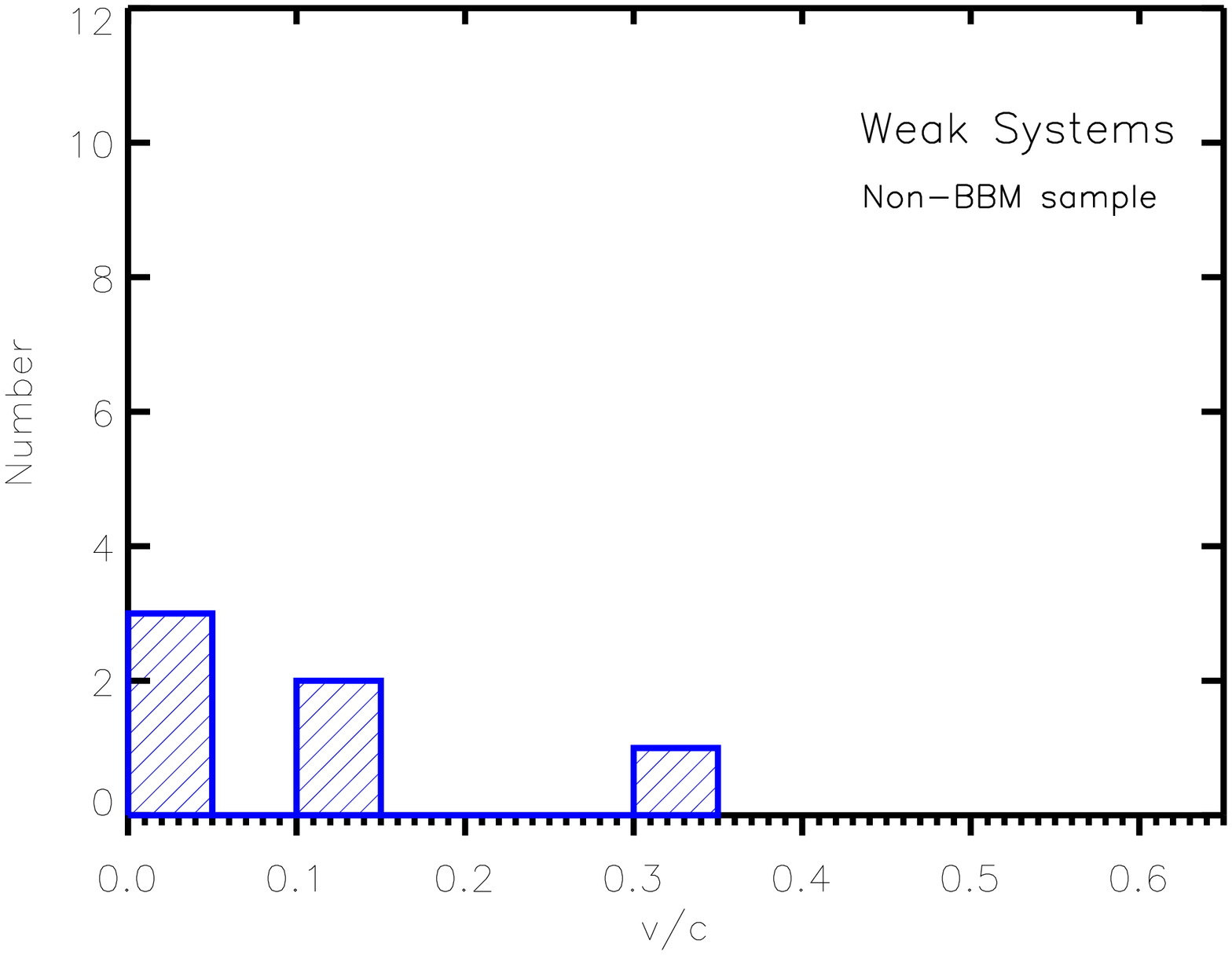}    
  \end{minipage}
 \hspace{0.8in}
  \begin{minipage}[]{0.4\textwidth}
  \includegraphics[width=0.8\textwidth,height=0.4\textheight,angle=90]{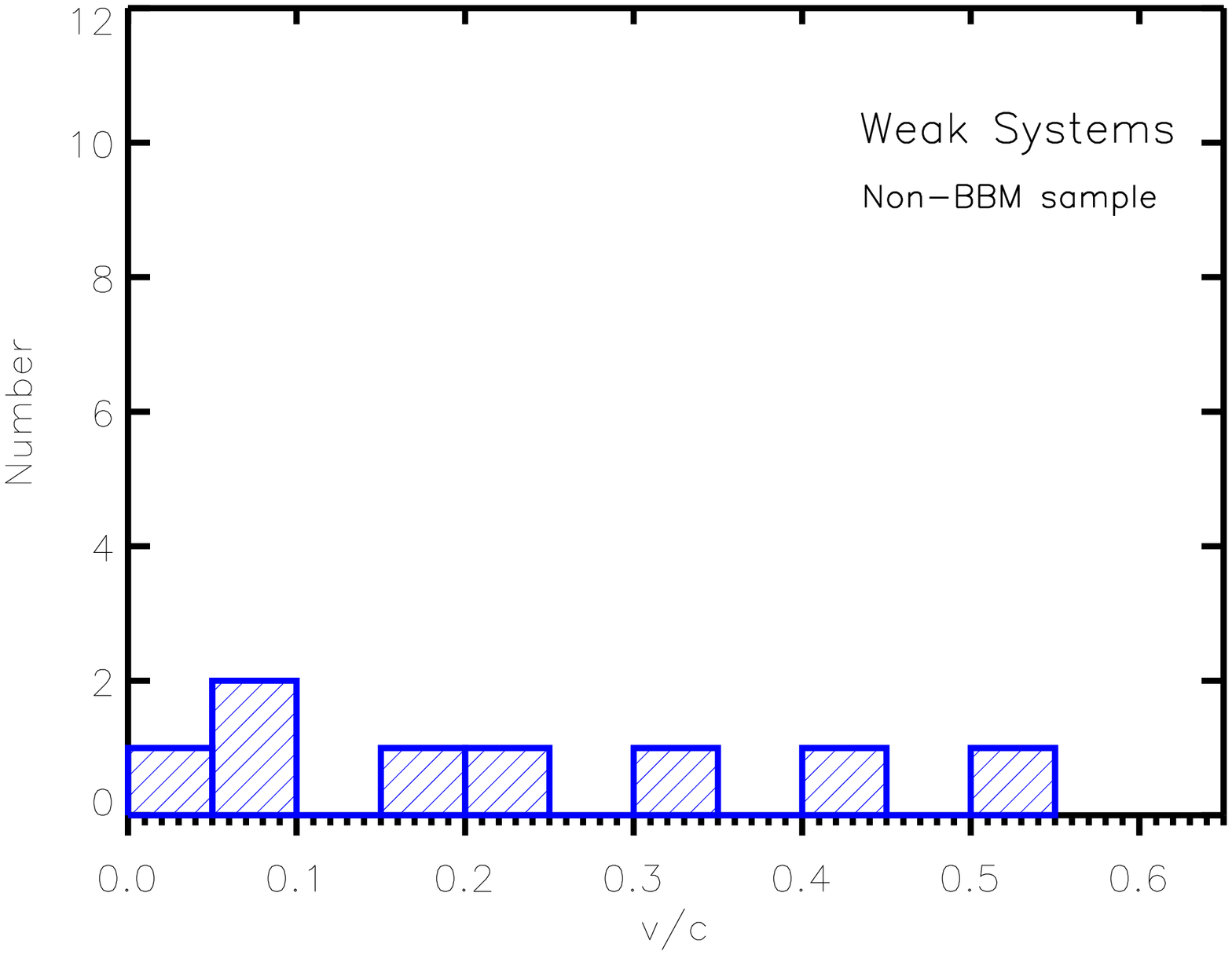}  
  \end{minipage}
   \caption[]{{\emph{Top:} Velocity distribution of the \mgii
       absorbers, relative to the background blazar (Eq.~\ref{eq:beta})
       for the 17 weak $(0.3~\AA~\le $W$_{\rm r}(2796) < 1.0 $~\AA~;
       \emph{left}) and 12 strong ( W$_{\rm r}(2796) \ge 1.0$~\AA~;
       \emph{right}) `intervening' Mg\,{\sc ii} absorbers seen towards
       the 42 BBM blazars, (see
       Section~\ref{subsection:reanalysis_bbm}).}
       {{\emph{Bottom:} The same
         as the top panel for the 7 weak and 8 strong `intervening'
         Mg\,{\sc ii} absorbers detected towards 160 blazars (among which only 64 have useful redshift path for weak systems)
         after excluding the BBM blazars.}}
     {Note that the absorption systems with offset velocities 
	  $\Delta v < 5000$ kms$^{-1}$, i.e., $\beta<0.017$ have been excluded.}
  }
\label{fig:beta_hist}
\end{figure*}

\begin{table*} 
\begin{minipage}[10]{130mm}
{\scriptsize
\caption{The results for the present sample and its various subsets, after excluding the \mgii absorption systems with $\Delta v< 60000$ kms$^{-1}$.}
\centering
\begin{tabular}{llllclllc}
\toprule 
Sample type$^{\dagger}$ & \multicolumn{4}{c}{Weak systems} & \multicolumn{4}{c}{Strong systems}\\\\
 & $N_{obs}$ & $\Delta$z & $\frac{N_{obs}}{\Delta z}$ & $\left(\frac{(N_{obs}/\Delta z)}{(dN/dz)_{qso}} \right)^{\alpha}$ 
 & $N_{obs}$ & $\Delta$z & $\frac{N_{obs}}{\Delta z}$ & $\left(\frac{(N_{obs}/\Delta z)}{(dN/dz)_{qso}} \right)^{\beta}$ \\\\
 \midrule 
Sample I     &  6&  18.80&$   0.32_{  0.13}^{  0.19}  $&$   0.76_{  0.30}^{  0.45}  $&   8&  39.68&$  0.20_{  0.07}^{  0.10}  $&$   1.10_{  0.38}^{  0.54}  $   \\\\                 
Sample II    &  5&  16.87&$   0.30_{  0.13}^{  0.20}  $&$   0.70_{  0.30}^{  0.47}  $&   7&  37.56&$  0.19_{  0.07}^{  0.10}  $&$   1.00_{  0.37}^{  0.54}  $   \\\\    
Sample III   &  5&  15.19&$   0.33_{  0.14}^{  0.22}  $&$   0.77_{  0.33}^{  0.52}  $&   7&  29.83&$  0.23_{  0.09}^{  0.13}  $&$   1.16_{  0.43}^{  0.62}  $   \\\\    	 
Sample IV    &  1&  7.38 &$   0.14_{  0.11}^{  0.31}  $&$   0.29_{  0.24}^{  0.67}  $&   4&  27.56&$  0.15_{  0.07}^{  0.11}  $&$   0.74_{  0.35}^{  0.58}  $   \\\\    	 
Sample V     &  1&  5.69 &$   0.18_{  0.15}^{  0.40}  $&$   0.36_{  0.30}^{  0.84}  $&   4&  19.83&$  0.20_{  0.10}^{  0.16}  $&$   0.85_{  0.41}^{  0.67}  $   \\\\    	 
Sample VI    &  1&  6.37 &$   0.16_{  0.13}^{  0.36}  $&$   0.33_{  0.27}^{  0.75}  $&   4&  25.12&$  0.16_{  0.08}^{  0.13}  $&$   0.79_{  0.38}^{  0.63}  $   \\\\    	 
Sample VII   &  1&  5.19 &$   0.19_{  0.16}^{  0.44}  $&$   0.39_{  0.32}^{  0.90}  $&   4&  18.19&$  0.22_{  0.11}^{  0.17}  $&$   0.91_{  0.44}^{  0.72}  $   \\\\
Sample VIII  &  4&  10.50&$   0.38_{  0.18}^{  0.30}  $&$   0.92_{  0.44}^{  0.73}  $&   3&  12.44&$  0.24_{  0.13}^{  0.23}  $&$   1.38_{  0.75}^{  1.35}  $   \\\\ 
Sample IX    &  4&  10.00&$   0.40_{  0.19}^{  0.32}  $&$   0.97_{  0.46}^{  0.77}  $&   3&  11.65&$  0.26_{  0.14}^{  0.25}  $&$   1.48_{  0.81}^{  1.44}  $   \\\\    	 
Sample X     &  5&  11.42&$   0.44_{  0.19}^{  0.30}  $&$   1.06_{  0.46}^{  0.72}  $&   4&  12.12&$  0.33_{  0.16}^{  0.26}  $&$   1.89_{  0.91}^{  1.50}  $  \\\\
\hline
\multicolumn{9}{l}{$^{\alpha}$ $dN/dz$ for QSOs is calculated at the mean value of the redshift path for the blazars, using equation 2 of BBM.} \\
\multicolumn{9}{l}{$^{\beta}$ $dN/dz$ for QSOs is calculated at the mean value of redshift path for the blazars, using equation 6 of BBM.}\\\\
\multicolumn{9}{l}{$^{\dagger}$The sample types are the same as in Table~\ref{Table:all_dndz_result} except that the regions with $\Delta v< 60000$ kms$^{-1}$ 
have been excluded here;}\\
\multicolumn{9}{l}{The additional Sample X in the last row corresponds to the 42 BBM blazars.}\\
  \end{tabular}
  \label{Table:all_dndz_result_60k}
}
\end{minipage}
\end{table*}

\section{Discussion and Conclusion}
\label{Section:Discussion_and_conclusions}
We have presented a new comparison of the incidence rates of \mgii
absorption systems towards two different classes of AGNs, namely
blazars and normal (optically selected) QSOs. A factor of two higher
rate towards blazars has earlier been claimed by BBM
(Section~\ref{sec:intro_mgiidndz}) and similar excess incidence of
intervening \mgii absorbers has been reported in a few earlier studies
of GRBs \citet{Prochter2006ApJ...648L..93P},
\citet{Sudilovsky2007ApJ...669..741S},
\citet{Vergani2009A&A...503..771V} and
\citet{Tejos2009ApJ...706.1309T}. However, the physical cause of the
purported excess relative to normal QSOs still remains to be
understood.  In fact, BBM have already discounted the possibilities of
dust obscuration and gravitational lensing playing a significant role
\citep[see also][]{Cucchiara2013ApJ...773...82C}.\par On the other
hand, no excess in the incidence rate of intervening \mgii absorbers
towards flat-spectrum radio quasars has been reported in some recent
studies based on large samples
\citep{Chand2012ApJ...754...38C,Joshi2013MNRAS.435..346J}.  Therefore,
in order to take a fresh look into the BBM finding of excess incidence
of \mgii absorbers along blazar sightlines, we have assembled a large
sample of 191 blazars (including the BBM sample of blazars).  An
independent sample of sightlines is also intended to provide a check
on the possible role of statistical fluctuation arising from small
source sample, as indeed turned out to be the case for GRBs
\citep[][Section~\ref{sec:intro_mgiidndz}]{Cucchiara2013ApJ...773...82C}\par
From the results of our analysis of the enlarged blazar sample
(Table~\ref{Table:all_dndz_result}), no excess is evident in the
$dN/dz$ along the sightlines to blazars, as compared to the sightlines
to normal QSOs. Recognizing that the spectral data for our blazar
sample have mostly rather modest SNR in comparison to the BBM sample
(see Fig.~\ref{fig:zemi_snr_dist}), we have sub-divided the spectra
for our blazar sample into two ranges of SNR (SNR between 5 and 15,
and SNR $>$ 15).  For neither of these SNR ranges did our analysis
show a significant difference between the $dN/dz$ estimates towards
blazars and normal QSOs (see Table~\ref{Table:all_dndz_result} and
Table~\ref{table:low_high_snr_comp}). Conceivably, the discrepancy
between our and BBM estimates of $dN/dz$ may then be rooted in the use
of different analysis procedures. However, this too seems unlikely
since our independent re-analysis of the BBM blazar sample reproduces
the $dN/dz$ excess reported by them
(Table~\ref{Table:bbm_excess_netzor}).\par To probe this issue
further, we compare in Fig.~\ref{fig:beta_hist}, the $\beta$
distributions of the \mgii absorbers for the BBM and our samples of
blazars. Here c$\beta$ is the velocity of an absorber measured
relative to the background blazar, where:
\begin{equation}
\beta \equiv \frac{v}{\rm c} = \frac {(1+z_{\rm em})^2-(1+z_{\rm
    abs})^2} {(1+z_{\rm em})^2+(1+z_{\rm abs})^2}
      \label{eq:beta}
\end{equation}
with, z$_{em}$ and z$_{abs}$ are the redshifts of the background AGN
and the \mgii absorber, respectively. The distributions shown in
Fig.\ref{fig:beta_hist} are useful for checking the extent of
clustering, if any, of \mgii absorbers up to mildly relativistic
$\beta$ values, as was noted in some recent studies of other AGN
samples (see below). The top two panels in Fig.~\ref{fig:beta_hist}
show the histograms of $\beta$ values of \mgii absorbers for the BBM
sample, both for weak (left panel of Fig.~\ref{fig:beta_hist}) and
strong absorbers (right panel in Fig.~\ref{fig:beta_hist}). The lower
two panels show the histograms for the \mgii absorbers for our blazar
sample, after excluding the BBM blazars. For the strong absorbers in
the BBM sample, a slight clustering at smaller $\beta$ is hinted,
which is consistent with the trend noticed in some earlier studies
\citep[][also BBM]{Joshi2013MNRAS.435..346J,Chand2012ApJ...754...38C},
as well as from Fig.~\ref{fig:cummulative_plot} (see above).  This
might indicate that associated \mgii absorbers may still contribute
significantly to $dN/dz$ up to offset velocities $\sim 0.2c$,
  especially for weak systems towards blazars (e.g., see
  Fig.~\ref{fig:cummulative_plot}, left panel).
Table~\ref{Table:all_dndz_result_60k} summarizes the $dN/dz$ estimates
for the various subsets of our 191 blazar sample, after excluding the
systems with $\Delta v< 60000$ kms$^{-1}$ i.e., $\beta < 0.2c$.\par
There is a hint of discrepancy when the \dndz excess for blazar
(relative to QSO sightlines) is compared for strong and weak
absorption systems, the excess being noticeable for weak absorbers
(e.g., see the top and middle panels in
Fig.~\ref{fig:cummulative_plot} ).  Attributing this marginally
significant excess to gas clumps accelerated outwards by the powerful
blazar jet, (e.g., up to $\Delta v< 60000$ kms$^{-1}$), as also noted
in BBM, the hinted excess in the case of weak absorbers could have its
origin in a physical cause, or merely an observational bias. For
instance, observationally the detection of gas clumps with higher
column density (i.e. stronger systems) would be easier as compared to
the lower column density clumps (i.e., weak absorbers). On the other
hand, occurrence of lower column density clumps is more likely,
intuitively. This seems to be the case as dynamical stability of the
relativistic jets suggests that external perturbations do not disrupt
the jets globally \citep[see, e.g.,][]{2017IAUS..324..141K}. This
means, in particular, that most of the clumps (or clouds) impinging on
the jet, as it propagates through the mostly diffuse gas, are smaller
than the jet radius. Assuming that clumps or clouds in the ambient
medium have similar volume densities, those with lower column
densities (hence weak systems) are likely to have a less disruption
effect on the jets via a slower growth of global instabilities. Hence,
lower column density clumps accelerated by the jets should be
intuitively more abundant in comparison to higher column density
clumps, consistent with the results shown in
(Fig.~\ref{fig:cummulative_plot}, left panel). The reality of the
discrepancy, however, remains to be confirmed using larger set of
blazar sighlines.  \par In summary, we conclude that (i) our
independent analysis of the spectral data used by BBM for their blazar
sample has reproduced the factor two excess claimed by them in $dN/dz$
for \mgii absorbers seen towards blazars, vis-a-vis normal QSO; (ii)
by using a $\sim 4$ times larger blazar sample (albeit, mostly with a
moderate SNR) which includes the BBM sample as well, we have arrived
at a statistically more robust and independent estimate of $dN/dz$ of
\mgii absorbers along blazar sightlines and the present analysis does
not show a significant difference from the $dN/dz$ known for the
sightlines to normal QSOs; (iii) the agreement improves further when
we limit the comparison to offset velocities above 60000
kms$^{-1}$. This would be consistent with the possibility that
associated \mgii absorbers remain a significant contributor to $dN/dz$
up to $\beta = 0.2c$ measured relative to the background QSO
\citep[see][also BBM]{Joshi2013MNRAS.435..346J}.
\par
Finally, in order to firmly settle the issues raised in the present
study a significant enlargement of the sample of \mgii absorbers
towards blazars would be vital. This can be achieved, e.g., by
extending the high sensitivity optical spectroscopic coverage to the
71 blazars which had to be excluded from the present analysis because
the SNR of their currently available spectra falls below our adopted
reasonable threshold (SNR $>$ 5).
\section*{Acknowledgments}
We thank the anonymous referee for his/her detailed comments to improve our manuscript.
 G-K thanks the National Academy of Sciences, India for the award of a
 Platinum Jubilee Senior Scientist fellowship. YS is supported by RFBR
 (17$-$52$-$45053$\_$IND).  We thank the scientific staff and the
 observing team at SAO for the help in our observations.  This
 research made use of (i) the Keck Observatory Archive (KOA), which is
 operated by the W. M. Keck Observatory and the NASA Exoplanet Science
 Institute (NExScI), under contract with the National Aeronautics and
 Space Administration, using observations made using the LRIS
 spectrograph at the Keck, Mauna Kea, HI; (ii) ESO Science Archive
 Facility by using observations made using the UVES, FORS and
 X-SHOOTER spectrographs at the VLT, Paranal, Chile.  Special thanks
 to John Pritchard from ESO user support for the useful discussions on
 the FORS pipeline. \par Funding for the SDSS and SDSS-II has been
 provided by the Alfred P.  Sloan Foundation, the Participating
 Institutions, the National Science Foundation, the U.S. Department of
 Energy, the National Aeronautics and Space Administration, the
 Japanese Monbukagakusho, the Max Planck Society, and the Higher
 Education Funding Council for England. The SDSS Web Site is
 http://www.sdss.org/. The SDSS is managed by the Astrophysical
 Research Consortium for the Participating Institutions. The
 Participating Institutions are the American Museum of Natural
 History, Astrophysical Institute Potsdam, University of Basel,
 University of Cambridge, Case Western Reserve University, University
 of Chicago, Drexel University, Fermilab, the Institute for Advanced
 Study, the Japan Participation Group, Johns Hopkins University, the
 Joint Institute for Nuclear Astrophysics, the Kavli Institute for
 Particle Astrophysics and Cosmology, the Korean Scientist Group, the
 Chinese Academy of Sciences (LAMOST), Los Alamos National Laboratory,
 the Max-Planck-Institute for Astronomy (MPIA), the
 Max-Planck-Institute for Astrophysics (MPA), New Mexico State
 University, Ohio State University, University of Pittsburgh,
 University of Portsmouth, Princeton University, the United States
 Naval Observatory, and the University of Washington.
\appendix
\section{Repersentative spectra of identified \mgii absorption systems}
\label{ap:appendix_A}
\begin{figure*}
  \includegraphics[width=1.0\textwidth,height=0.8\textheight]{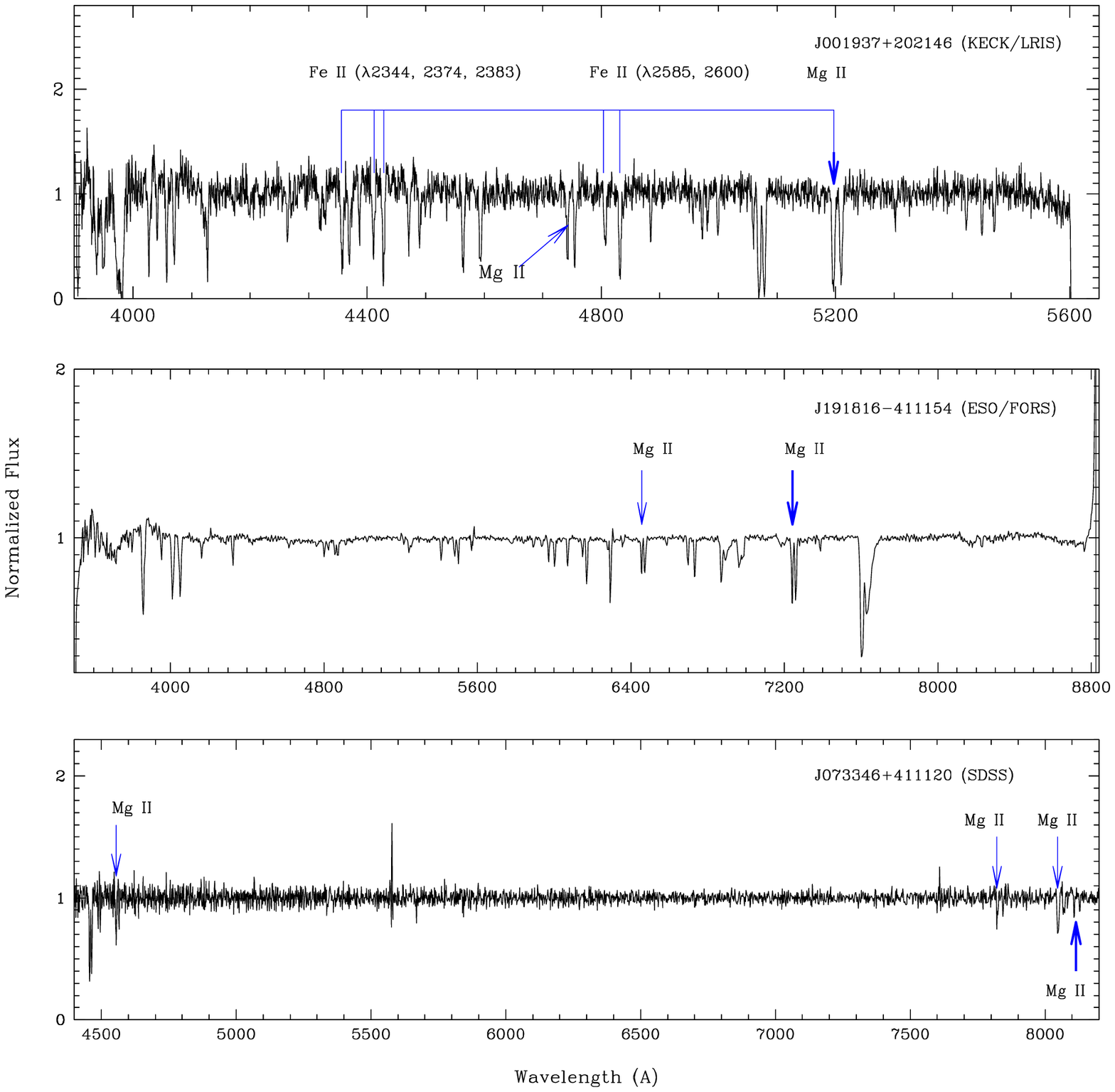}
  \caption{Representative normalized blazar spectra: J211552$+$000115 
  (\emph{top}),
    J191816-411154 (\emph{middle}), J073346+411120(\emph{bottom}). In 
	each case, the most redshifted \mgii absorption system (thick blue arrow) detected was used as the lower limit to the blazar's emission redshift.  
	Thin blue lines in \emph{top panel} mark the
    \feii($\lambda$2344,2374,2383,2585,2600) lines associated with the
    \mgii absorption line system at z$_{\rm abs}=0.85840$.}
  \label{fig:asso_system_spectra}	
\end{figure*}

For 51 out of the 191 blazars that form our sample, emission redshift 
measurement are not available and therefore we have taken lower limit
estimates based on the most redshifted Mg II absorber seen in their spectra.
A representative normalized spectrum is shown in 
Fig.~\ref{fig:asso_system_spectra} for 3 of these 51 blazars, with brief 
comments as below.\par

As seen from Fig.~\ref{fig:asso_system_spectra} (top panel) for
J001937$+$202146, two \mgii absorption systems were identified, (i) at
$\lambda = 4743.06$ \AA\ (z$_{\rm abs}=0.69616$,
W$_{r}$(\mgii)=1.0969 \AA\ ) (ii) at $\lambda = 5196.74$\AA\ (z$_{\rm
  abs}=0.85840$, W$_{r}$(\mgii)=2.7312 \AA\ ). For both these systems,
corresponding absorptions lines due to \feii($\lambda$2344,2374,2383,2585,2600) 
and \mgi($\lambda$2852) are detected.\par

In J191816$-$411154 (Fig.~\ref{fig:asso_system_spectra} middle panel),
two \mgii absorption doublets were detected (i) at $\lambda =
6457.92$ \AA\ (z$_{\rm abs}= 1.30941$, W$_{r}$(\mgii) = 0.9761 \AA\ )
together with the corresponding \feii($\lambda$2344, 2374, 2383, 2586, 2600)
and \mgi($\lambda$2852) absorption lines; and (ii) at $\lambda
= 7243.02$ (z$_{\rm abs}= 1.59017$, W$_{r}$(\mgii)= 1.5710 \AA\ ) together 
with the corresponding absorption lines due to \feii($\lambda$2344, 2374, 
2383, 2586, 2600) and \mgi($\lambda$2852), \aliii($\lambda\lambda$1854,1862) 
and \siiv($\lambda\lambda$1393,1402).  
The redshift of the more redshifted \mgii absorption system was used as the 
lower limit of the emission redshift for this source.\par

The blazar J073346$+$411120 (Fig.~\ref{fig:asso_system_spectra} bottom
panel), has four \mgii absorption systems: (i)  at $\lambda =
4555.11$ \AA\ (z$_{\rm abs}= 0.62895$, W$_{r}$(\mgii) = 0.72 \AA\ ),
(ii) at $\lambda =7821.67 $ \AA\ ($z_{\rm abs}= 1.79710$,
W$_{r}$(\mgii)=0.27 \AA\ ), (iii) at $\lambda = 8046.36$
\AA\ ($z_{\rm abs}= 1.87745$, W$_{r}$(\mgii) = 0.59 \AA\ ) and (iv)  at
$\lambda =8107.74 $ \AA\ ($z_{\rm abs}= 1.89940$, W$_{r}$(\mgii) = 0.07
\AA\ ), with weak corresponding \feii($\lambda$2383) absorption feature
detected for all four systems.  The absorption redshift of the most 
redshifted \mgii absorption system (i.e., $z_{\rm abs}= 1.89940$) was used 
as the lower
limit of the emission redshift for this source. The systems at $z_{\rm
abs}= 1.87745$ could, however, not be included in the analysis as it falls
within 5000 kms$^{-1}$ of the lower limit to the source's emission redshift, 
which we have fixed at $z_{\rm abs}= 1.89940$).
\bibliography{references}
\label{lastpage}
\end{document}